\documentclass[aps,prd,twocolumn,floats,balancelastpage,showpacs,preprintnumbers,floatfix,nofootinbib,superscriptaddress]{revtex4}
 
\usepackage{graphicx}
\usepackage{graphics}
\usepackage{amsmath}
\usepackage{amssymb,amsfonts}
\usepackage{footnote}

\newcommand{\beq}{\begin{equation}}
\newcommand{\eeq}{\end{equation}}
\newcommand{\ben}{\begin{eqnarray}}
\newcommand{\een}{\end{eqnarray}}
\newcommand{\bi}{\begin{itemize}}
\newcommand{\ei}{\end{itemize}}

\newcommand{\sv}{\ensuremath{\langle\sigma_{\text{ann}}v\rangle}}

\begin{document}

\title{Search for dark matter annihilation signatures in H.E.S.S. observations of Dwarf Spheroidal Galaxies}

\author{A.~Abramowski}
\affiliation{Universit{\"a}t Hamburg, Institut f{\"u}r Experimentalphysik, Luruper Chaussee 149, D 22761 Hamburg, Germany}

\author{F.~Aharonian}
\affiliation{Max-Planck-Institut f{\"u}r Kernphysik, P.O. Box 103980, D 69029 Heidelberg, Germany}
\affiliation{Dublin Institute for Advanced Studies, 31 Fitzwilliam Place, Dublin 2, Ireland}
\affiliation{National Academy of Sciences of the Republic of Armenia, Marshall Baghramian Avenue, 24, 0019 Yerevan, Republic of Armenia}

\author{F.~Ait Benkhali}
\affiliation{Max-Planck-Institut f{\"u}r Kernphysik, P.O. Box 103980, D 69029 Heidelberg, Germany}

\author{A.G.~Akhperjanian}
\affiliation{National Academy of Sciences of the Republic of Armenia, Marshall Baghramian Avenue, 24, 0019 Yerevan, Republic of Armenia}
\affiliation{Yerevan Physics Institute, 2 Alikhanian Brothers St., 375036 Yerevan, Armenia}

\author{E.~Ang{\"u}ner}
\affiliation{Institut f{\"u}r Physik, Humboldt-Universit{\"a}t zu Berlin, Newtonstr. 15, D 12489 Berlin, Germany}

\author{M.~Backes}
\affiliation{University of Namibia, Department of Physics, Private Bag 13301, Windhoek, Namibia}

\author{S.~Balenderan}
\affiliation{University of Durham, Department of Physics, South Road, Durham DH1 3LE, U.K.}

\author{A.~Balzer}
\affiliation{GRAPPA, Anton Pannekoek Institute for Astronomy, University of Amsterdam,  Science Park 904, 1098 XH Amsterdam, The Netherlands}

\author{A.~Barnacka}
\affiliation{Obserwatorium Astronomiczne, Uniwersytet Jagiello{\'n}ski, ul. Orla 171, 30-244 Krak{\'o}w, Poland}
\affiliation{and now at Harvard-Smithsonian Center for Astrophysics,  60 Garden St, MS-20, Cambridge, MA 02138, USA}

\author{Y.~Becherini}
\affiliation{Department of Physics and Electrical Engineering, Linnaeus University,  351 95 V\"axj\"o, Sweden}

% ...

\collaboration{The H.E.S.S. Collaboration}
\noaffiliation

\begin{abstract}
Dwarf spheroidal galaxies of the Local Group are close satellites of the Milky Way characterized by a large mass-to-light ratio and are not expected to be the site of non-thermal high-energy gamma-ray emission or intense star formation. Therefore they are amongst the most promising candidates for indirect dark matter searches. During the last years the High Energy Stereoscopic System (H.E.S.S.) of imaging atmospheric Cherenkov telescopes observed five of these dwarf galaxies for more than 140 hours in total, searching for TeV gamma-ray emission from annihilation of dark matter particles. The new results of the deep exposure of the Sagittarius dwarf spheroidal galaxy, the first observations of the Coma Berenices and Fornax dwarves and the re-analysis of two more dwarf spheroidal galaxies already published by the H.E.S.S. Collaboration, Carina and Sculptor, are presented. In the absence of a significant signal new constraints on the annihilation cross-section applicable to Weakly Interacting Massive Particles (WIMPs) are derived by combining the observations of the five dwarf galaxies. The combined exclusion limit depends on the WIMP mass and the best constraint is reached at 1--2 TeV masses with a cross-section upper bound of $\sim$ 3.9$\times$10$^{-24}$ cm$^3$ s$^{-1}$ at a $95\%$ confidence level.
\end{abstract}

\maketitle

\section{Introduction}

A large number of observations from Galactic to cosmological scales support the hypothesis that dark matter (DM) should be primarily composed of a new type of particle of yet unknown nature. A popular class of candidates are stable or very long-lived weakly interacting massive particles (WIMPs). With masses and couplings falling roughly within the electroweak scale, they are predicted by numerous theories beyond the Standard Model of particle physics and could account for the total amount of DM inferred from the thermal relic picture~\cite{bib:[2]}. WIMP searches mostly follow three types of strategies: searches at colliders, and notably the Large Hadron Collider (LHC), probing both the production of DM particles themselves and other signatures of extensions of the Standard Model that could be of relevance to DM physics; searches for nuclear recoil signals in direct detection experiments, probing the WIMP scattering cross-section on ordinary matter; and finally indirect searches for a signal in the products of potential WIMP annihilations in astrophysical observations, probing the corresponding cross-section.

DM-induced gamma rays can present sharp spectral signatures, like for instance $\gamma \gamma$ or $Z \gamma$ annihilation lines, with energy trivially related to the WIMP mass. However, since DM is electrically neutral, these processes are loop-suppressed and therefore typically very rare. WIMP-induced gamma rays are thus expected to be dominated by a relatively featureless continuum of byproducts of cascades and decays (mostly from $\pi^0\to \gamma\gamma$) following the annihilation in pairs of quarks, gauge/Higgs bosons or leptons. The number of resulting gamma rays depends quadratically on the DM density along the line of sight of the observer. This motivates a number of promising targets for indirect DM searches, namely those with expected DM density enhancements against conventional astrophysical processes, in particular the Galactic Centre, galaxy clusters and nearby dwarf spheroidal galaxies. 

This paper presents the final results in constraining DM annihilation in five dwarf spheroidal galaxies (dSph) observed with the H.E.S.S. experiment during its first phase, conducted with four telescopes with $13$~m diameter mirror. In particular, new results are obtained with a more sensitive analysis of the Sagittarius dSph for which the H.E.S.S. collaboration conducted a deep exposure over the past 6 years. In addition the paper presents constraints for the previously not observed Coma Berenices and Fornax dSphs, as well as a combination of all five dwarf galaxies observed with the H.E.S.S. four telescope configuration, including two more dSphs previously observed with H.E.S.S. namely Carina and Sculptor.

\section{Dwarf spheroidal galaxies}

The dSphs of the Local Group are believed to be amongst the best targets to search for gamma-ray signals from the annihilation of DM particles and to derive robust constraints on the annihilation cross-section\cite{Lake:1990du,Stoehr:2003hf,Evans:2003sc}. Indeed, these satellites of the Milky Way (MW) are located at ${\cal O}$(100~kpc) and are essentially free of gamma-ray background since they are characterized by properties such as little to no gas, dust or recent star formation. Their mass-to-luminosity ratios are as high as a few hundreds, amongst the highest in the Universe (see~\cite{Aaronson,Mateo:1998wg} and references therein). As discussed in section~\ref{hierarchicalmassmodel}, the dynamical study of the stellar component embedded in their DM halos, facilitates the reduction of the uncertainties concerning the spatial distribution of DM particles in these systems (for a review see~\cite{battaglia}).\\

\begin{center}
{\it Sagittarius dSph} 
\end{center}
The Sagittarius dwarf spheroidal galaxy (Sgr), discovered in 1994, is the nearest dwarf galaxy to the MW, at a distance of ˜25~kpc~\cite{bib:[4]}. According to photometric measurements, its nominal position is spatially coincident with the globular cluster M54~\cite{bib:[17]}: RA = 18$^{\mathrm{h}}$ 55$^{\mathrm{m}}$ 03$^{\mathrm{s}}$, Dec = $-30^{\circ}$ 28$^{\mathrm{\prime}}$ 42$^{\mathrm{\prime}\mathrm{\prime}}$ in equatorial coordinates (J2000.0). Sgr changed its orbits over its lifetime substantially~\cite{beloku14} and it was severely influenced by Galactic tides, pulling large numbers of stars from the core to form stellar streams that wrap around the Galaxy at least once and contributing to the build-up of the MW stellar halo system~\cite{deboer14}. Although the stellar kinematics in the Sgr remnant indicates the presence of DM, even if not at the same high density levels observed in other dSphs, an ambiguity on its mass exists since the structure, size and origin of the Sgr progenitor are very uncertain. Some recent works~\cite{nieder11} inferred a progenitor luminosity of the same order of the present-day Small Magellanic Cloud, modelling Sgr with a mass-to-luminosity ratio lower than the other dSphs. Furthermore some works~\cite{law} have also provided ample evidence that Sgr has an unusual large number of associated globular clusters when compared to the typical dwarf spheroidal galaxies. Recent measurements of its stellar kinematics however, strongly resemble those of other classical dSphs~\cite{Frinchaboy}, indicating that Sgr may not be an outlier anomaly, but rather a tidally disrupted, DM-dominated system like several other satellites of the MW.  

On the motivation that it is subject to significant tidal stripping which could affect the determination of the integrated DM density, bounds from Sgr are often derived under ultra-conservative assumptions~\cite{Essig:2009jx} or dropped altogether (as in the recent Fermi-LAT analysis~\cite{Ackermann:2013yva}). 

In the present study however, the uncertainties of the integrated astrophysical factor are included as nuisance parameters in the likelihood profile. The impact of removing Sgr from the stacked data set is discussed in section~\ref{subsec:likelihood}.\\

\begin{center}
{\it Coma Berenices dSph}
\end{center}
The Coma Berenices dSph was recently discovered in the Sloan Digital Sky Survey~\cite{COM_2} and is located at a distance of about 44~kpc, centered at RA = 12$^{\mathrm{h}}$ 26$^{\mathrm{m}}$ 59$^{\mathrm{s}}$, Dec = 23$^{\circ}$ 54$^{\mathrm{\prime}}$ 15$^{\mathrm{\prime}\mathrm{\prime}}$ in equatorial coordinates (J2000.0). Coma Berenices is one of the smallest and faintest satellites of the MW, having extreme low luminosities, differing from the average characteristics of other dSphs in the plane of absolute magnitude $vs$ half-light radius~\cite{COM_2}. However, spectroscopic surveys reveal kinematics and metallicities in line with those of dwarf galaxies. Coma Berenices, claimed to be amongst the most DM dominated dSphs~\cite{strigari}, is fairly regular in shape and does not show important signs of tidal debris according to a recent deep, wide-field photometric survey~\cite{munozComa}.\\

\begin{center}
{\it Fornax dSph}
\end{center}
The Fornax dSph is a well established satellite of the MW~\cite{Shapley,Mateo:1998wg}, located at a distance of about 140~kpc, RA = 2$^{\mathrm{h}}$ 39$^{\mathrm{m}}$ 59.3$^{\mathrm{s}}$, Dec = $-34^{\circ}$ 26$^{\mathrm{\prime}}$ 57$^{\mathrm{\prime}\mathrm{\prime}}$ in equatorial coordinates (J2000.0). The stellar kinematical data for Fornax suggest that it is DM dominated with a mass-to-light ratio of order 20 within its optical extent. Fornax hosts five globular clusters and other substructures~\cite{goerdt} and the observed stellar kinematics are dominated by random motions without evidence of tidal disruption~\cite{walker}.\\

\begin{center}
{\it Carina and Sculptor dSphs}
\end{center}
Carina~\cite{Mateo:1998wg,Munoz:2006hx} and Sculptor~\cite{Westfall:2005ji} dSphs complete the list of targets considered in this work. Carina dSph is located at a distance of 101~kpc, with equatorial coordinates (J2000.0): RA = 06$^{\mathrm{h}}$ 41$^{\mathrm{m}}$ 36$^{\mathrm{s}}$, Dec = $-50^{\circ}$ 57$^{\mathrm{\prime}}$ 58$^{\mathrm{\prime}\mathrm{\prime}}$. Sculptor dSph is closer, 79~kpc, with coordinates: RA = 01$^{\mathrm{h}}$ 00$^{\mathrm{m}}$ 09$^{\mathrm{s}}$, Dec = $-33^{\circ}$ 42$^{\mathrm{\prime}}$ 32$^{\mathrm{\prime}\mathrm{\prime}}$ (equatorial J2000.0). These two targets are amongst the most luminous dSphs near the MW. The best estimates of the orbits of the two dSphs show that Carina is likely to be more tidally disrupted than Sculptor, leading to higher uncertainties for the DM content of the Carina dSph than of the Sculptor. However, the extent of the disruption in Carina remains matter of controversy: precise measurements of the direction of its proper motion does not support a tidal origin for its elongation, while the difference in the position angles is not significant enough to rule out such an origin~\cite{piatekCarina}. The first constraints from the search for a DM signal in these two targets have been reported in~\cite{hessScCa}. 
  

\section{H.E.S.S. observations and analysis}

The High Energy Stereoscopic System (H.E.S.S.) is an array of Imaging Atmospheric Cherenkov Telescopes (IACTs) designed to study high energy gamma-ray emitters by recording the faint Cherenkov light induced by air showers in the atmosphere. Located in the Khomas Highland in Namibia, 1800 m above sea level, H.E.S.S. began operation in 2004 with four 13\,m telescopes equipped with cameras containing 960 photo-multiplier tubes. The H.E.S.S. array operates in coincidence mode with at least two telescopes triggered within a coincidence window of 60~ns for an event to be accepted.

\subsection{Observations and data selection}
\label{sec:datapreparation}

The five dSphs data sets were acquired from 2006 to 2012, during different observation campaigns and with different total exposure time. The observations were performed in \textit{wobble mode}, where the source is offset from the centre of the field of view, enabling simultaneous background estimation.  
All data have been calibrated following the standard calibration and selection procedures~\cite{aha06b}. Further quality cuts have been applied. The most relevant requirements are a minimum number of three telescopes in operation during the observation runs, a central trigger rate after zenith angle correction between 100 and 400~Hz, a fraction of inactive pixels per camera and per observation run, not larger than 15\%, the reconstructed image of a camera should contain at least a charge amplitude of 60 photo-electrons and, to avoid truncated images which might lead to misreconstructed events, shower images with centre of gravity reconstructed at more than 2$^{\circ}$ degrees from the centre of the camera are neglected.
In the following the data sets, the analysis procedures and the results of the analysis of the complete dwarf spheroidal galaxies data set are described and discussed.

\subsubsection*{Sagittarius dSph}
Following a first observation campaign dedicated to Sgr in 2006, which collected 11 hours of useful data, and in absence of any signal, the H.E.S.S. collaboration published an upper limit on the gamma-ray flux and a constraint on the velocity-weighted annihilation cross-section for DM annihilation~\cite{bib:para}. The H.E.S.S. collaboration continued to observe Sgr using the four 13\,m telescopes from 2007 to 2012, accumulating 90 hours of quality selected data. The complete Sgr data sample was taken with the source at different zenith angles, spanning from a few degrees up to 45$^{\circ}$ with an average value of about 16$^{\circ}$. 

\subsubsection*{Coma Berenices dSph}
The Coma Berenices dSph was observed with the H.E.S.S. experiment from 2010 to 2013. The total effective live-time after quality selection corresponds to 8.6 hours. The observations were performed with relatively high zenith angles with a mean value of about 48$^{\circ}$.

\subsubsection*{Fornax dSph}
The Fornax dSph was observed with larger camera offsets since it was not the primary target of the corresponding data set. The analysis results of about 6.0 hours of data acquisition are reported in this work. The average zenith angle of the observations is about 14$^{\circ}$.

\subsubsection*{Carina and Sculptor dSphs}
The Carina and Sculptor dSphs were observed with the H.E.S.S. experiment between 2008 and 2009. The data set consists of a total of 12.7 and 12.5 hours, respectively. In absence of any signal from their nominal positions the H.E.S.S. collaboration published upper limits on the gamma-ray flux as well as constraints on the velocity-weighted annihilation cross-section of DM particles~\cite{hessScCa}. The same published data set from the Sculptor observations is considered in this work, while the Carina data set comprises 10 additional hours, acquired more recently. The mean zenith angle of the Carina dSph data set is about 35$^{\circ}$, and 14$^{\circ}$ for Sculptor dSph.

\subsection{Data analysis}

The $X_{\rm eff}$ analysis~\cite{xeff} was employed for the selection of gamma-ray events and for the suppression of cosmic-ray background events. The $X_{\rm eff}$ method improves the separation of gamma-ray and cosmic-ray events compared to the standard H.E.S.S. analysis~\cite{aha06b} by exploiting the complementary discriminating variables of three reconstruction methods used in H.E.S.S. and usually referred to as Hillas~\cite{hillas}, Model~\cite{model} and 3D-model~\cite{model3D, godo}. The resulting unique discriminating variable, called $X_{\rm eff}$, acts as an event-by-event gamma-misidentification probability estimator, combining the probability density functions for events identified as gamma-ray-like or hadron-like by the three reconstruction methods. The final gamma-ray-like event selection was achieved through a set of cuts adapted to the detection of faint sources~\cite{xeff}\footnote{Specifically, the values of the cuts employed are $\eta=0.7$ and $X_{\rm eff,cut}=0.3$}. For the energy and direction of each reconstructed candidate event the values provided by the Model method were selected. The search for gamma-ray signals was conducted assuming point-like sources translating into an angular size cut of $\theta \leq 0.1^{\circ}$.


\begin{figure}[ht]
  \centering
  \includegraphics[width=0.3\textwidth]{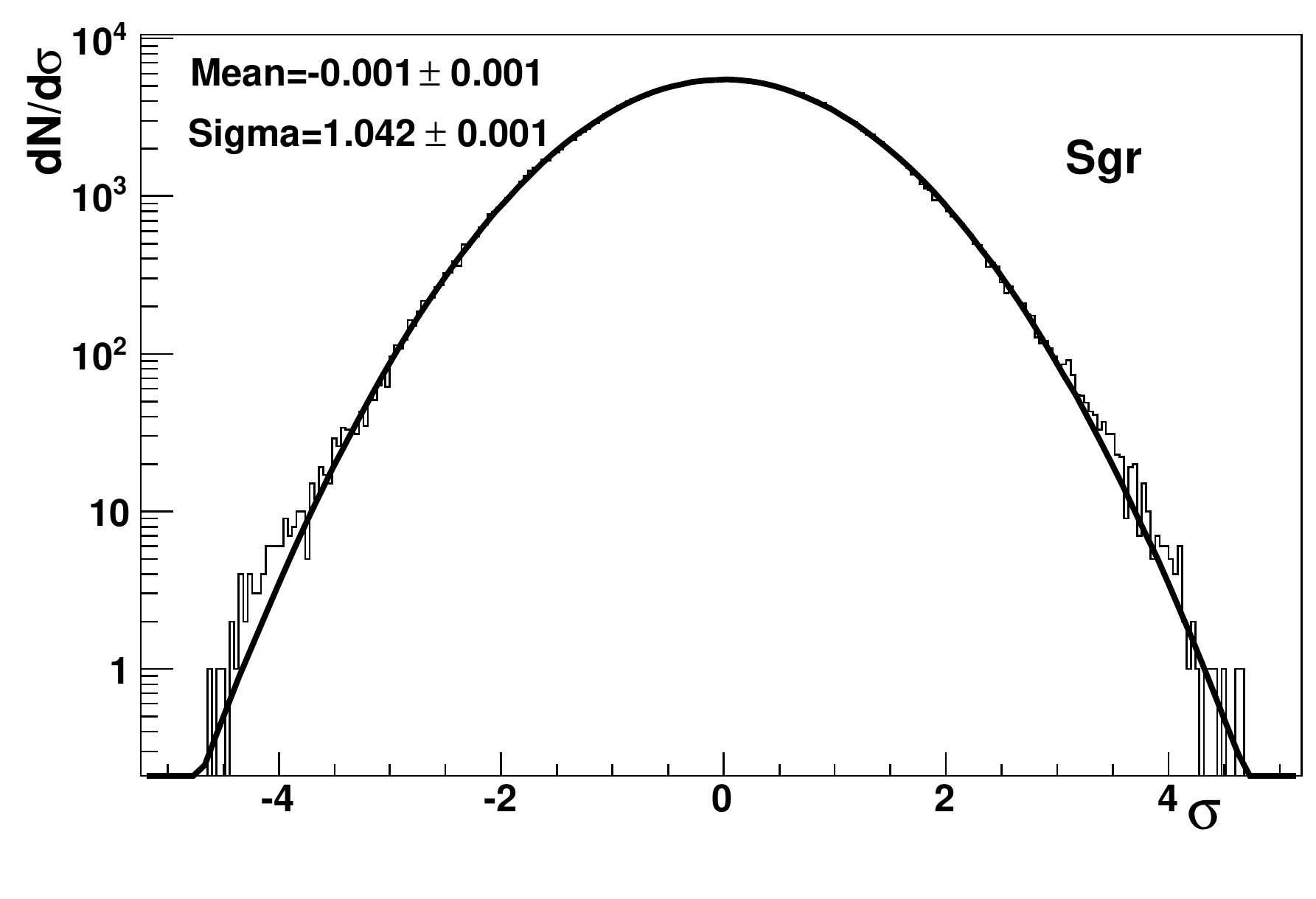}
  \includegraphics[width=0.3\textwidth]{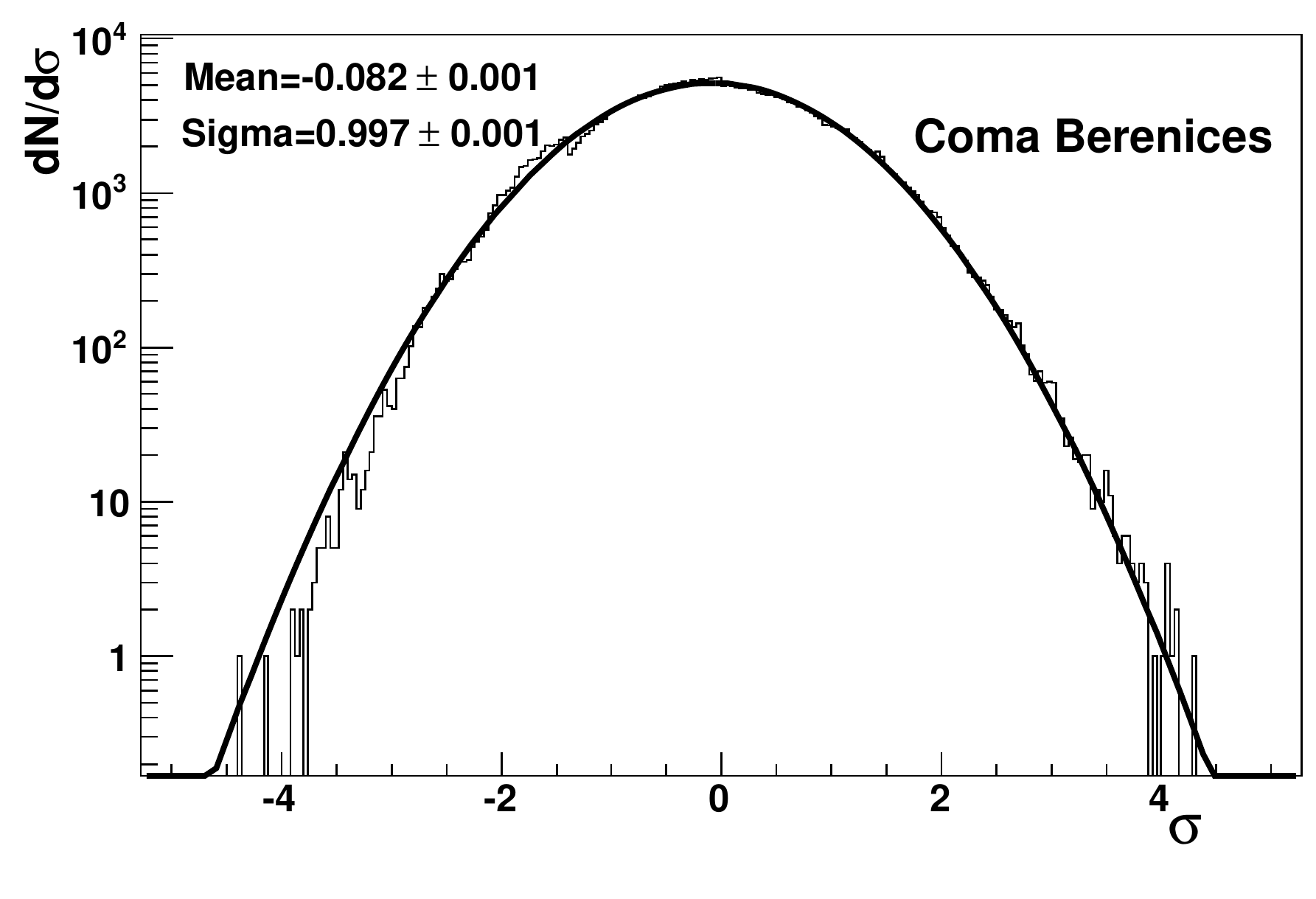}
  \includegraphics[width=0.3\textwidth]{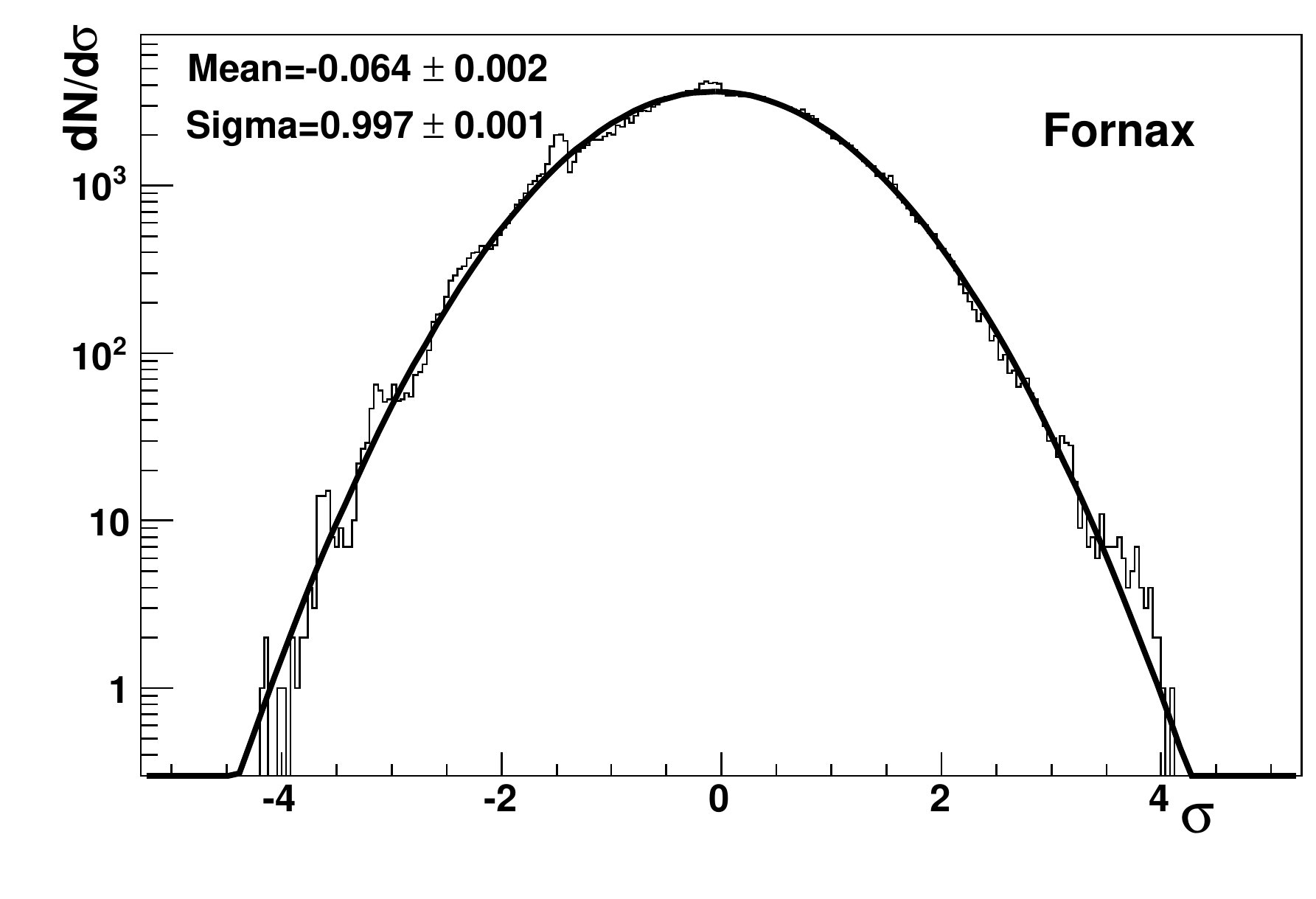}
  \includegraphics[width=0.3\textwidth]{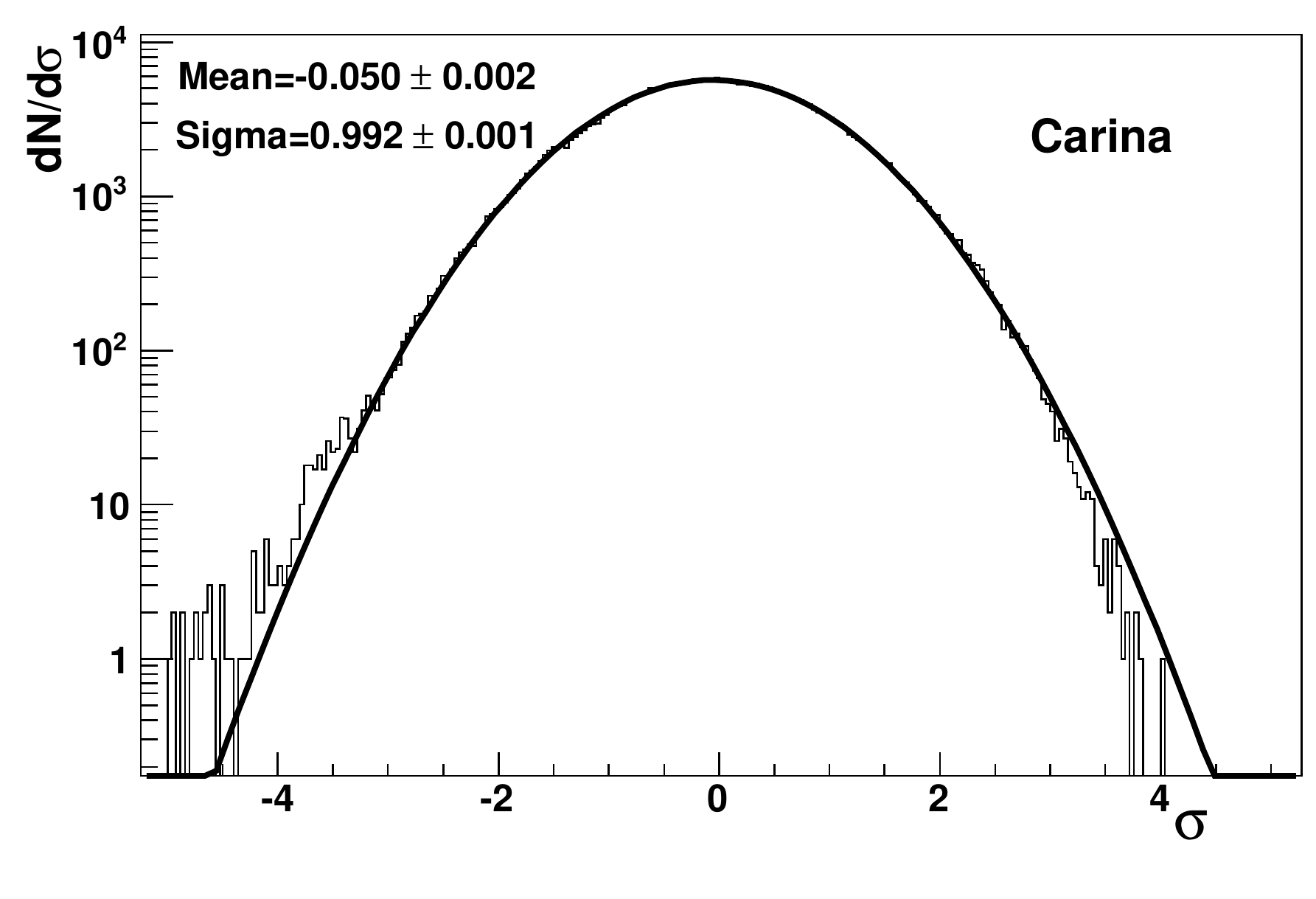}
  \includegraphics[width=0.3\textwidth]{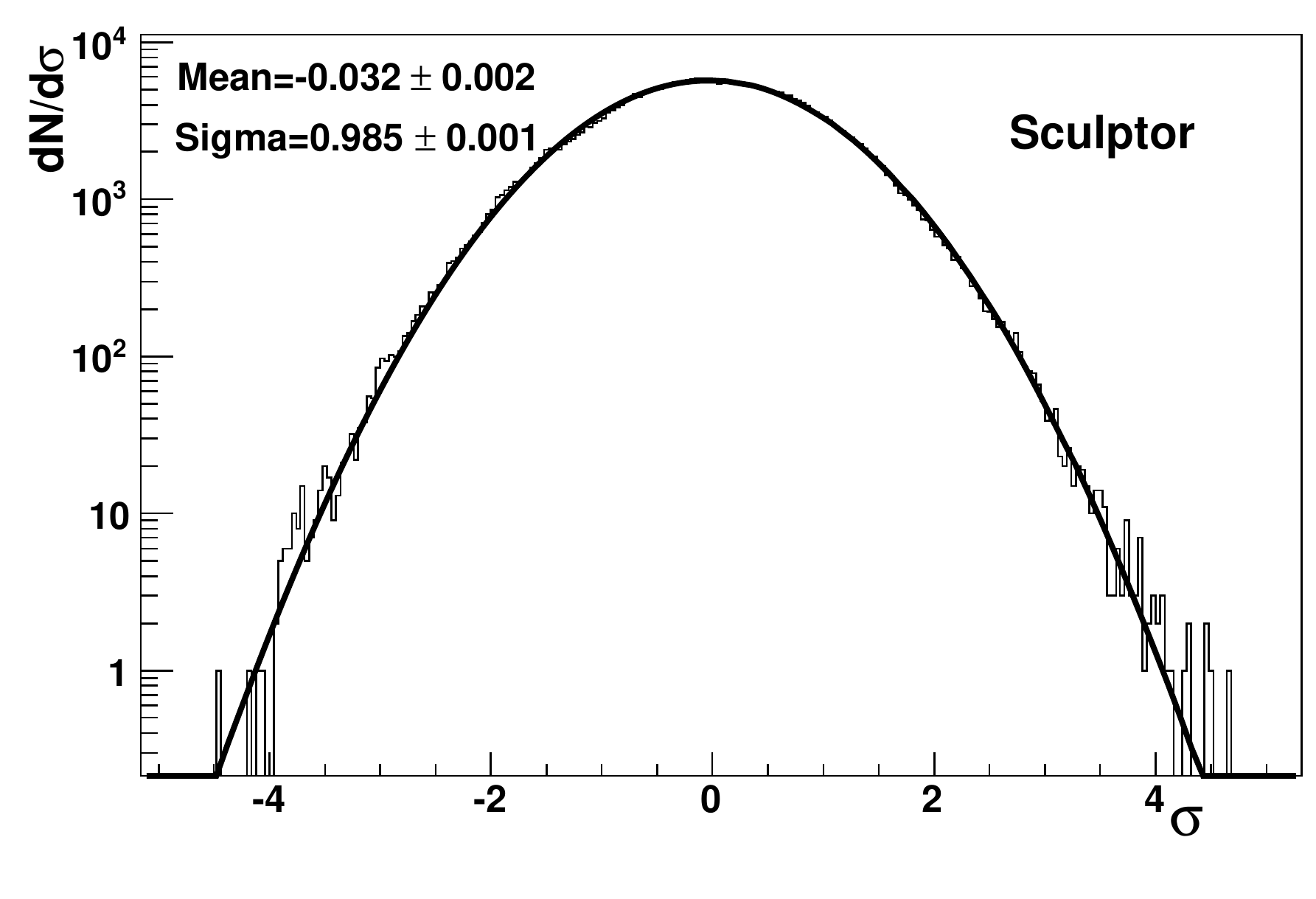}
  \caption{Significance distributions and corresponding Gaussian fits over the camera field of view for the five analyzed dSphs. No significant excess is seen at the nominal target positions.}
  \label{fig:sigmas}
\end{figure}

\begin{figure*}[ht]
  \centering
  \includegraphics[width=0.35\textwidth]{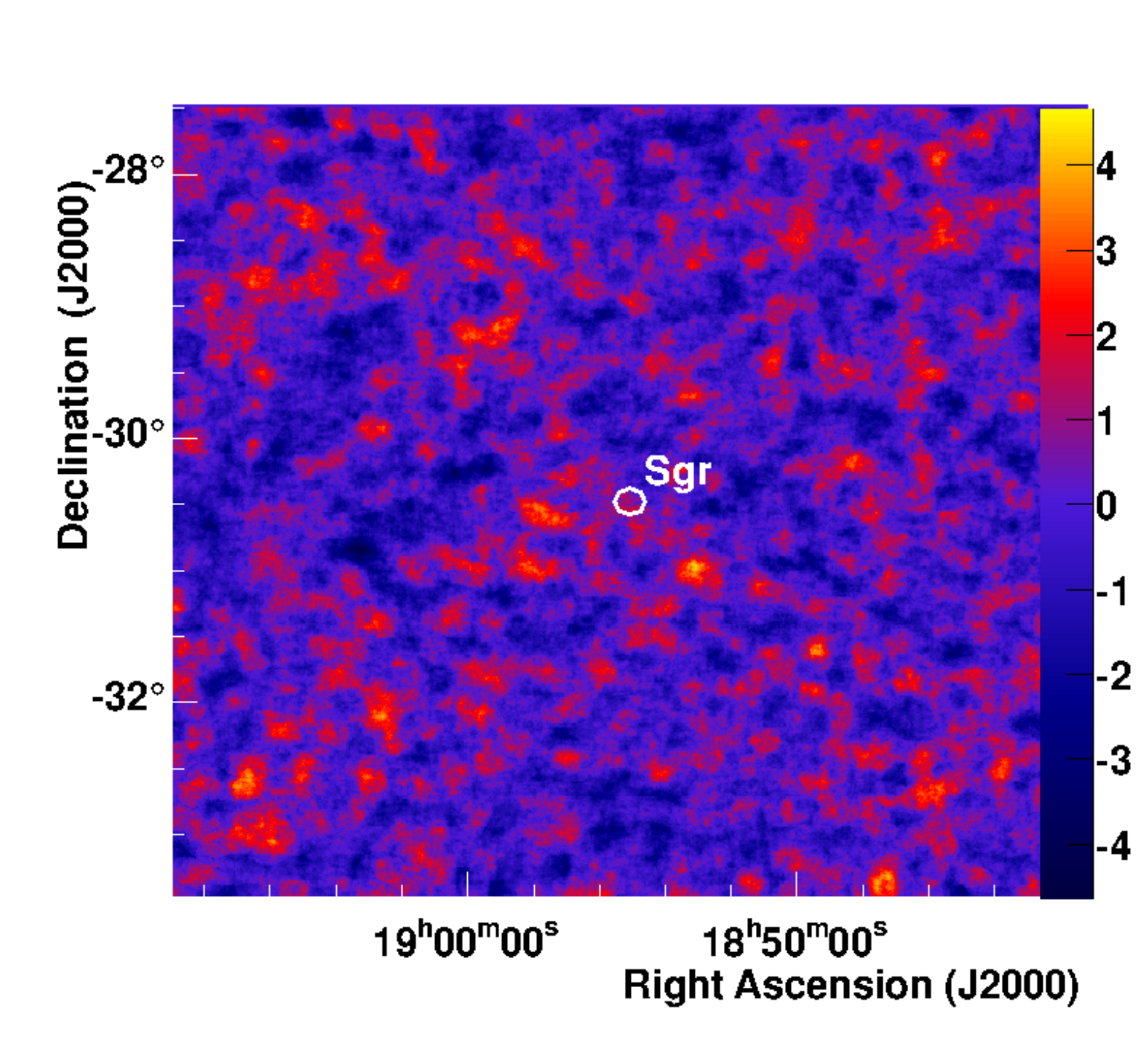}
  \includegraphics[width=0.46\textwidth]{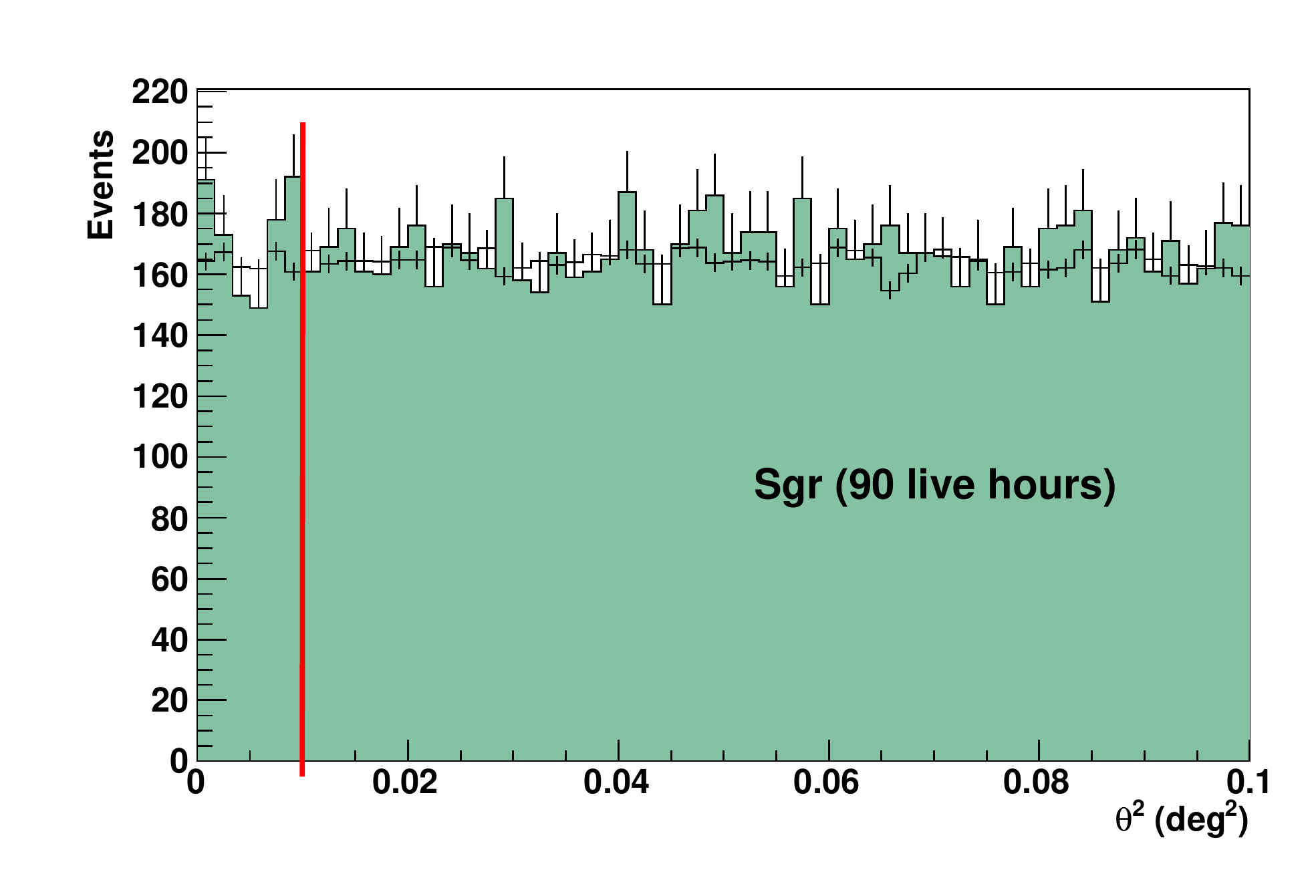}
  \caption{\textit{Left:} significance skymap in equatorial coordinates. \textit{Right:} $\theta^2$ radial distribution of the ON events for gamma-ray-like events from the Sgr target position. The estimated background is also shown by black crosses. No significant excess is seen within an angular region around the source position of $\theta \leq 0.1^{\circ}$.}
  \label{fig:theta}
\end{figure*}

\subsection{Results}
All dwarf spheroidal galaxies have been analyzed with the same procedure. Using the \emph{ring background} technique~\cite{Berge2007}, no significant deviations from the estimated backgrounds have been found at the nominal positions of the five dSphs or in the target field of view. The distribution of significances from the dSphs fields of view are well compatible with a Gaussian profile centered on zero, as shown in Figure~\ref{fig:sigmas}.
This can also be seen in Figure~\ref{fig:theta}, where the squared angular distributions ($\theta^2$) of events from source (\textit{ON}) and control (\textit{OFF}) regions obtained from the analysis of Sgr are shown to be compatible. For the five dSphs the resulting significances range from $-1.8\,\sigma$ to 2.7$\,\sigma$. From the number of events registered in the source, $N_{\rm on}$, and those corresponding from the control regions, $N_{\rm off}$, and their exposure ratio, $\alpha$, the 95\% confidence level (C.L.) upper limit on the total number of observed gamma-ray events , $N_{\gamma}^{95\% C.L.}$,  was computed using the Rolke et al. method~\cite{bib:rolke} for each source.

All data analysis results together with the observation conditions specific to each dSph are summarized in Table~\ref{analy}. For completeness the results obtained reanalyzing Sculptor and Carina dSphs data are also shown; they are compatible with the ones previously published in~\citep{hessScCa}.  

\section{Dark matter flux}

The differential gamma-ray flux ($d\Phi_\gamma/dE_\gamma$) due to DM particle annihilations depends on the following factors:
\begin{itemize}
\item the particle physics (both Standard Model and beyond) processes involved in the DM annihilation;
\item the DM density distribution at the source (hereafter referred to as ``halo profile'');
\item the solid angle $\Delta\Omega$ within which the signal is integrated along the line of sight of the observer;
\item the interstellar gas density and radiation density, concerning the gamma rays induced by DM-produced secondary $e^\pm$ losing energy by inverse Compton scattering and relativistic bremsstrahlung. Usually, the resulting radiation falls well below the energy threshold of Cherenkov telescopes \cite{Colafrancesco:2006he} and it will be ignored in the following. Here only primary gamma-ray emission from $\pi^0 \to \gamma \gamma$ will be considered.
\end{itemize}

\noindent
The differential flux is hence usually factorized as:
\begin{equation}
\frac{d\Phi_\gamma}{dE_\gamma}(E_\gamma, \Delta\Omega)  =  \Phi^{pp}(E_\gamma) \times J(\Delta\Omega)\Delta\Omega,
\end{equation}
where the first factor ($\Phi^{pp}$) encodes information on the underlying particle physics DM model, while the second factor (hereafter referred to as $J$-factor) depends on the astrophysical DM density distribution at the source. 
\\
\\
The particle physics factor can be written as:
\begin{equation}
\Phi^{pp} = \frac{d\Phi_\gamma}{dE_\gamma}  =  \frac{1}{8\pi} \frac{\langle\sigma_{\rm ann}v\rangle}{m^2_\chi}  \times  \frac{dN_\gamma}{dE_\gamma},
\end{equation}
where $\langle\sigma_{\rm ann}v\rangle$ is the total velocity-averaged self-annihilation cross-section, $m_{\chi}$ is the WIMP particle mass, and $dN_\gamma/dE_\gamma$ is the differential gamma-ray spectrum per WIMP annihilation.

\begin{savenotes}
\begin{table*}[!t]
  \caption{Summary of the observation conditions and data analysis results per each dSph: the average observational zenith angle; the average minimum energy threshold ($E_\mathrm{th}$); the acceptance corrected livetime; the number of events detected in the target region $N_\mathrm{on}$; the acceptance corrected exposure ratio $\alpha$; the number of events detected in control regions $N_\mathrm{off}$; the resulting significance $\sigma$; the 95\% C.L. upper limit on the total number of observed gamma-ray events $N_{\gamma}^{95\% C.L.}$ and, if the resulting significance is negative, the average upper limit at 95\% C.L., obtained with the expected background and no true excess signal assumed, is also quoted.}
  \label{analy}
  \centering
  \begin{tabular}{c c c c c c c c c c }
\\
  \hline
  \hline
  
dSph & Mean Zenith ($^{\circ}$) & $E_\mathrm{th} (\mathrm{GeV}$) & Live-time (hrs) & $N_\mathrm{on}$ & $\alpha^{-1}$ & $N_\mathrm{off}$ & $\sigma$ & $N_{\gamma}^{95\% C.L.}$ \\
   \hline
   \hline
   Sagittarius & 15.99 & 196 & 90.0 & 820 & 17.94 & 13652 & 2.05  & 117.8\\
Coma Berenices & 47.75 & 714 & 8.6  & 25  & 12.99 & 459   & -1.78 & 5.8 (14.0)\\
        Fornax & 13.90 & 292 & 6.1  & 24  & 49.30 & 648   & 2.65  & 21.8\\
        Carina & 35.36 & 356 & 23.2 & 108 & 17.00 & 2031  & -1.03 & 13.0 (24.5)\\
      Sculptor & 14.21 & 264 & 12.5 & 96  & 19.28 & 1909  & -0.30 & 18.2 (22.4) \\
  \hline
   \end{tabular}
  \end{table*}
\end{savenotes}

The velocity-averaged annihilation cross-section $\langle\sigma_{\rm ann}v\rangle$ can be computed within the framework of each specific particle physics model providing a DM candidate. A rough estimate for its magnitude in {\it thermal} prediction scenarios is given by the well-known value $\langle\sigma_{\rm ann}v\rangle \sim 3\times10^{-26}$ cm$^3$ s$^{-1}$ \cite{Jungman:1995df}.

The differential gamma-ray spectrum per WIMP annihilation depends on the composition of the primary DM annihilation products. It can be written as
\begin{equation}
\frac{dN_\gamma}{dE_\gamma} = \sum_i B_i \frac{dN^i_\gamma}{dE_\gamma}
\end{equation}
where $B_i$ and $dN^i_\gamma/dE_\gamma$ are the branching fractions into the $i$-th final state and its respective gamma-ray yield. The composition of the final state particles, which can either be Standard Model particles or more exotic states, is also model-dependent and constitutes one of the major particle physics uncertainties.

\subsection{The particle physics factor}

In the case of massive gauge or Higgs bosons, quark and $\tau$ final states, gamma rays are produced mainly through the decay and hadronization of the annihilation products. In recent years, significant effort has been devoted to provide more accurate calculations of the gamma-ray yield of such final state particles \cite{Cembranos:2010dm,Cirelli:2010xx}, taking advantage of the respective evolution of Monte Carlo event generators~\cite{Sjostrand:2006za,Sjostrand:2007gs,Corcella:2000bw}. In this work, the results presented in \cite{Cembranos:2010dm} are utilised in order to estimate the gamma-ray flux per DM annihilation at the source for a few annihilation channels. In particular, the authors provide fitting functions of $dN^i_\gamma/dx$, where $x = E_\gamma/m_{\chi}$, for different DM mass ranges, the maximal mass value being $8$ TeV. In the following analysis, when considering DM particle masses $m_{\chi} > 8$ TeV the fitting function corresponding to $m_{\chi} = 8$ TeV will be employed. It should be kept in mind that results for heavy DM masses are anyway only indicative, since significant corrections are expected to the tree-level two-body final state contribution typically included in theoretical cross-sections calculations~\cite{Ciafaloni:2010ti,Bringmann:2013oja}.

When the final state consists of light leptons, the gamma-ray production mechanism differs from the one described previously and is dominated by final state radiation (FSR) of photons with, in the case of muons, an additional contribution arising from radiative muon decay into electrons. The authors of \cite{Cembranos:2010dm} provide results for both the $e^+ e^-$ and the $\mu^+ \mu^-$ final states, which are used in this analysis. In a variety of models however, e.g. in leptophilic models of heavy DM with light mediators $\phi$ resulting in a large Sommerfeld enhancement of $\langle\sigma_{\rm ann}v\rangle$, DM can annihilate into a four-lepton final state through light mediator pair production as $\chi\chi\rightarrow\phi\phi \rightarrow\ l^+ l^- l^+ l^-$. In this case, the gamma-ray spectrum per DM annihilation is estimated analytically in the following way: 

\begin{itemize}
 \item In the case of a four-electron final state, gamma rays arise as a result of FSR from the final state leptons. This effect can be adequately described in the mediator $\phi$ rest frame by the Weizs{\"a}cker-Williams approximation and is given, for collinear photon emission, by the Altarelli-Parisi splitting function. The total spectrum can then be calculated by performing a Lorentz boost of the resulting gamma-ray distribution to the DM rest frame, which essentially coincides with the interstellar medium rest frame.
 \item In the case of a four-muon final state, in addition to the above FSR contribution, there is also the spectrum coming from radiative muon decays, see e.g.~\cite{Kuno:1999jp}. It can induce a non-negligible contribution to the total gamma-ray spectrum which, for mediator masses $m_\phi$ of ${\cal O}(1 \mathrm{GeV})$ and DM masses of ${\cal O}(10 \mathrm{TeV})$, can be as large as $20-30\%$~\citep{Essig:2009jx}. A first Lorentz boost allows one to go to the mediator rest frame, a second one to the DM rest frame and compute the relevant spectrum.
\end{itemize}
For the sake of brevity, the relevant analytical expressions have been omitted. A fairly concise description can be found in \cite{Essig:2009jx}.

In what follows the following expression for the gamma-ray yield (taken from~\cite{bib:berg}) will be moreover considered:
\begin{equation}
\label{Berg}
\frac{dN_\gamma}{dE_\gamma}  =  \frac{1}{ m_{\chi} } \frac{ dN_\gamma }{dx} = 
\begin{cases}
\frac{1}{ m_{\chi} } \frac{0.73 e^{-7.8x}}{x^{1.5}}, & \text{if\,} x \leq 1\\
0, & \text{otherwise}
\end{cases}.
\end{equation}
\noindent
This parametrisation tries to capture a representative supersymmetric spectrum for neutralino DM annihilating in $W^+W^-$ and $Z\,Z$ final state spectra. This expression will be used in addition to the other single-particle final states in order to facilitate comparison with previous studies. The final annihilation channel in $b\bar{b}$, parametrized according to~\cite{Cembranos:2010dm}, is also considered. The different physical mechanisms translate into qualitatively distinct gamma-ray spectra.  Final states composed of massive gauge or Higgs bosons and quarks yield gamma rays through complex processes of hadronization and decay of the annihilation products. This produces relatively ``soft'' spectra albeit characterised by a relatively large number of photons. FSR emission instead is relatively hard, but a higher order process in perturbation theory and corresponds to a significantly reduced normalisation in the number of photons. Moreover, for kinematic reasons, the gamma-ray spectrum from a four-lepton final state is softer than that from a two-lepton final state, and the centre of mass energy associated to the lepton production process is the moderate one of the mediator mass, reducing further the photon yield.  Finally, since $\tau$ leptons possess both hadronic and leptonic decay modes with comparable branching ratios, the $\tau \tau$ channel corresponds to an intermediate situation between the two extreme spectra mentioned above.

\subsection{The astrophysical factor}
\label{hierarchicalmassmodel}
The expected differential flux of gamma rays from DM particle annihilation depends also on the astrophysical factor $J$. This factor is defined as the integral along the line of sight of the squared density of the DM distribution in the observed object and it is averaged over the solid angle of the observation, as
\begin{equation}
J = \frac{1}{\Delta\Omega}\iint_{\Delta\Omega} \rho^2_{\rm DM}(l,\Omega)dld\Omega.
\label{j}
\end{equation}
\noindent In this work a solid angle $\Delta\Omega$ = 10$^{-5}$ sr is considered, consistent with the point-spread function of the instrument as achieved in the analysis~\cite{xeff}, since a signal from an almost point-like source is sought for.

For the five dwarf galaxies considered, the DM mass densities were derived following~\citep{Martinez2013}. In this work, a Bayesian two-level likelihood analysis is performed, enabling to simultaneously constrain the properties of individual dSphs of the local group, as well as those of the entire MW satellites population. 
The bottom-level describes the astrophysical properties of each individual dSph and its underlying DM potential: the total set of observables are the line-of-sight velocities, metallicites, and positions of individual stars in the galaxy, as well as the total galaxy luminosity. The top-level describes the overall distribution of halo properties. The total model parameter set is composed of the stellar profile, DM profile, and stellar velocity anisotropy parameters.

Many questions in galaxy formation are affected by limited knowledge of the stellar velocity dispersion anisotropy and the limited ability to quantify the amount of DM in the outer parts of elliptical galaxies. It has been shown that for each dispersion-supported galaxy, there exists one radius, the (3D) stellar half-light radius, within which the integrated mass as inferred from the line-of-sight velocity dispersion is largely insensitive to stellar velocity dispersion anisotropy. Within this radius the mass is well characterised by a simple formula that depends only on quantities that may be inferred from observations~\cite{wolf2}. The prior probability which constraints the bottom-level observables is based on the basic assumption that the satellite galaxies share this underlying property. Using this approximation for each dSph, the bottom-level data set is consequently composed of the mass enclosed within the half-light radius, the measured half-light radius, the total luminosity and their associated errors. Further prior assumptions are that: the enclosed mass is dominated by the DM contribution; a log-log profile-independent relationship is applied between the maximum circular velocity and radius corresponding to this velocity~\citep{Martinez2013}. Finally the underlying DM density profiles $\rho_{\rm DM}$ are parametrised through any model. These sets of data are then in turn constrained by the top-level likelihood. 

It has proved hard to distinguish the structure of the DM halo, especially at small radii. The available observational evidence for Carina~\cite{gilmore}, Fornax and Sculptor~\cite{walker2009} tended to suggest that the DM density was shallower than the 1/r density cusp, while observed velocity dispersion profiles and cored light distributions were likely to be consistent with DM halos with both central cores and cusps. 
Some recent works~\cite{amorisco2011,battaglia2008,walkpena} have modelled separately the velocity dispersion profiles of different metal-poor and metal-rich star populations contained in the dSphs. They all conclude that the data tend to privilege cored DM haloes than cusped one.
For the five dSphs examined in this analysis, two different distributions of DM particles in dSphs, a NFW~\cite{bib:[15]} and a Burkert~\cite{burkert} profile, are considered. While the first presents a cusp structure in the central halo region, the latter corresponds to an empirical form of the DM particle distribution, based on observed rotation curves of dSphs and larger galaxies, which resembles an isothermal distribution with a central core.


The results of the described two-level method for all dSphs considered in this study, except for Sgr, are already provided in~\citep{Martinez2013}. For Sgr the bottom-level data set was obtained from the analysis of the line-of-sight velocities of stars taken from~\cite{Frinchaboy}. 
This sample contains both Sgr stars and foreground stars from our Galaxy from 24 separate fields, representing  the most extended survey conducted on Sgr so far. Using the Besan{\c c}on model~\citep{BesanconModel}, the MW foreground distribution of stars was modeled for the 24 samples. For each field, a fit was performed with a model of velocity dispersion of stars belonging to both the MW and Sgr. First, using the mass estimator for dispersion-supported stellar systems of~\cite{wolf2}, the mass at half-light radius was inferred from the derived velocity dispersion. Then, a correction accounting for triaxiality was performed, according to the results of simulations reported in~\cite{Kowalczyk}. The total set of posteriors resulting from the first-level likelihood, as well as the variables parametrizing the priors, were employed as inputs to the second level likelihood.

{
\renewcommand{\arraystretch}{1.5}
\begin{table}[ht]
\caption{Mode and 1$\,\sigma$ uncertainty of the $J$-factor lognormal posterior, assuming an integration solid angle $\Delta\Omega = 10^{-5}$~sr and according to the mass distribution expectation values computed following~\cite{Martinez2013}.\label{tab:dwarfs}}
\centering
\begin{tabular}{|c|c|c|} 
\hline
\hline & \multicolumn{2}{c|}{$\mathrm{log}_{10}\left( \frac{J}{\mathrm{GeV}^2 \mathrm{cm}^{-5}} \right)$} \\ 
 dSph & NFW & Burkert \\
\hline
\hline
Sagittarius    & $19.1 \pm 0.5$ & $18.5 \pm 0.5$ \\
Coma Berenices & $18.8 \pm 0.4$ & $19.1 \pm 0.2$ \\
Fornax         & $18.1 \pm 0.3$ & $18.4 \pm 0.3$ \\
Carina         & $18.0 \pm 0.4$ & $18.4 \pm 0.2$ \\
Sculptor       & $18.5 \pm 0.3$ & $18.8 \pm 0.2$ \\
\hline
\end{tabular}
\end{table}
}

The resulting expectation values of $J$-factors together with their corresponding uncertainties are listed in Table~\ref{tab:dwarfs}. The J factors derived assuming Burkert profile are marginally consistent with those derived assuming an NFW profile. With the exception of Sgr, the mode of the $J$-factor posteriors derived with the Burkert profile are slightly larger than for the NFW modeling. This can be understood since two competing effects enter the determination of the $J$-factor, the density of the profile and the extent of volume integrated over, compared to the characteristic scale radius of the profile. The putative DM signal for most dSphs is basically point-like: Burkert profile fits lead to comparatively higher densities at larger radii, which dominate the $J$-factor provided they are included in the volume integral (i.e. they are within the detector point spread function, PSF). However, if the angular extent of the source is somewhat larger than the PSF (as qualitatively expected for more massive and nearby objects) a relatively higher contribution to the signal comes from the inner region, hence cuspier profiles like NFW lead to larger $J$-factors than shallower ones.

A recent detailed modeling of Sgr DM halo using an evolutionary N-body simulation of Sgr within the MW was obtained in the case of an isothermal DM profile, taking into account the tidal disruption of the dSph~\cite{bib:[16]}. In this study, an additional truncation of the DM halo at a radius of $\simeq$ 4~kpc was introduced to account for the tidal disruption observed in this system. The corresponding $J$-factor found is of the order of $10^{18}\mathrm{GeV}^2 \mathrm{cm}^{-5}$, with an uncertainty of more than a factor of two~\cite{bib:[16]}. The value derived in case of the Burkert profile using the multi-level likelihood technique is fully compatible with their finding.

\begin{figure}[t]
  \centering  \includegraphics[width=0.45\textwidth]{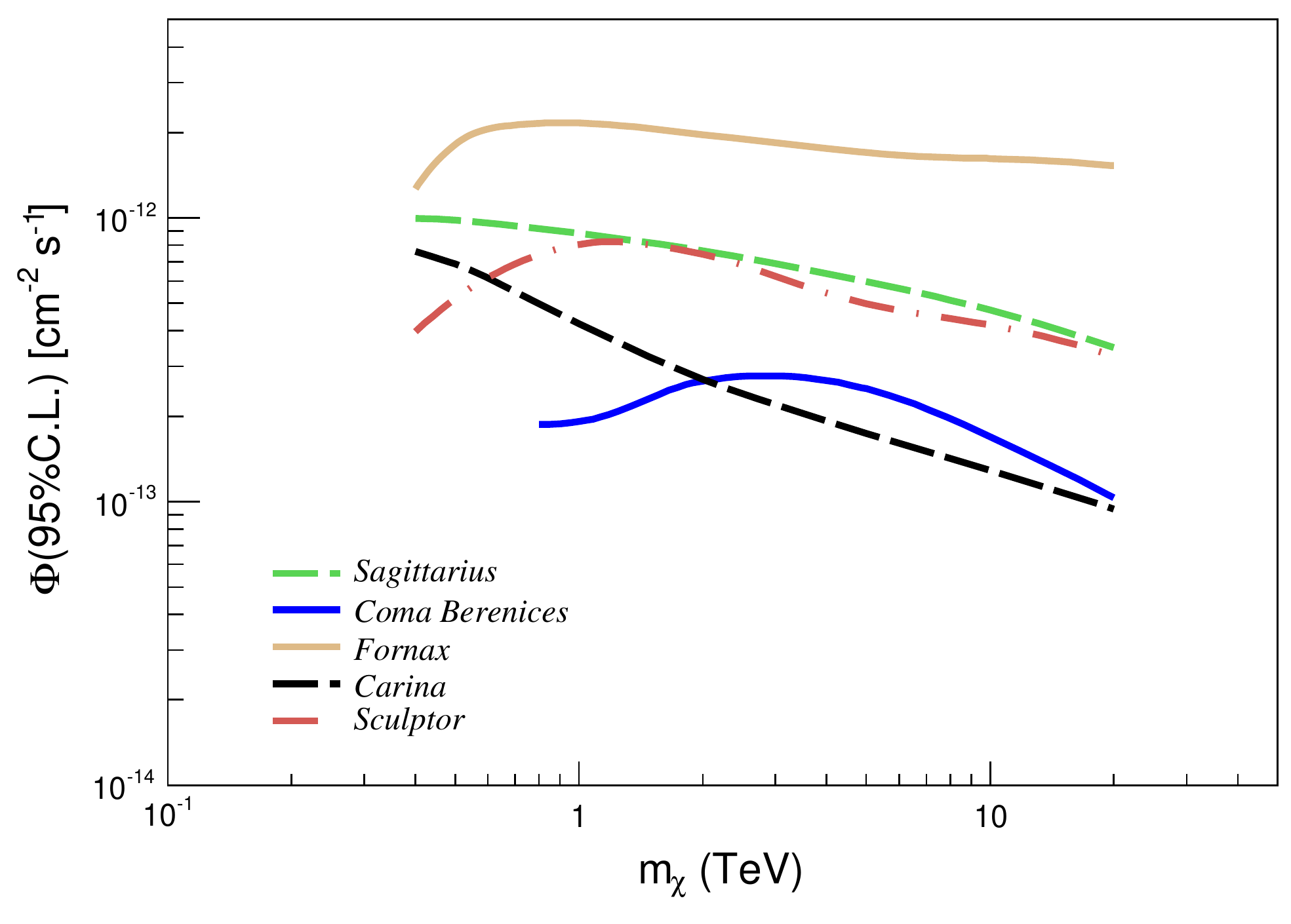}
  \caption{Upper limits at 95\% C.L. on the flux as a function of $m_{\chi}$ and under the hypothesis of DM particle annihilation in the $W^+W^-$ and $Z\,Z$ final states as parametrized in Eq.~(\ref{Berg}), obtained with the likelihood approach.}
  \label{ulFlux}
 \end{figure}
\begin{figure}[t]
  \centering
\includegraphics[width=0.45\textwidth]{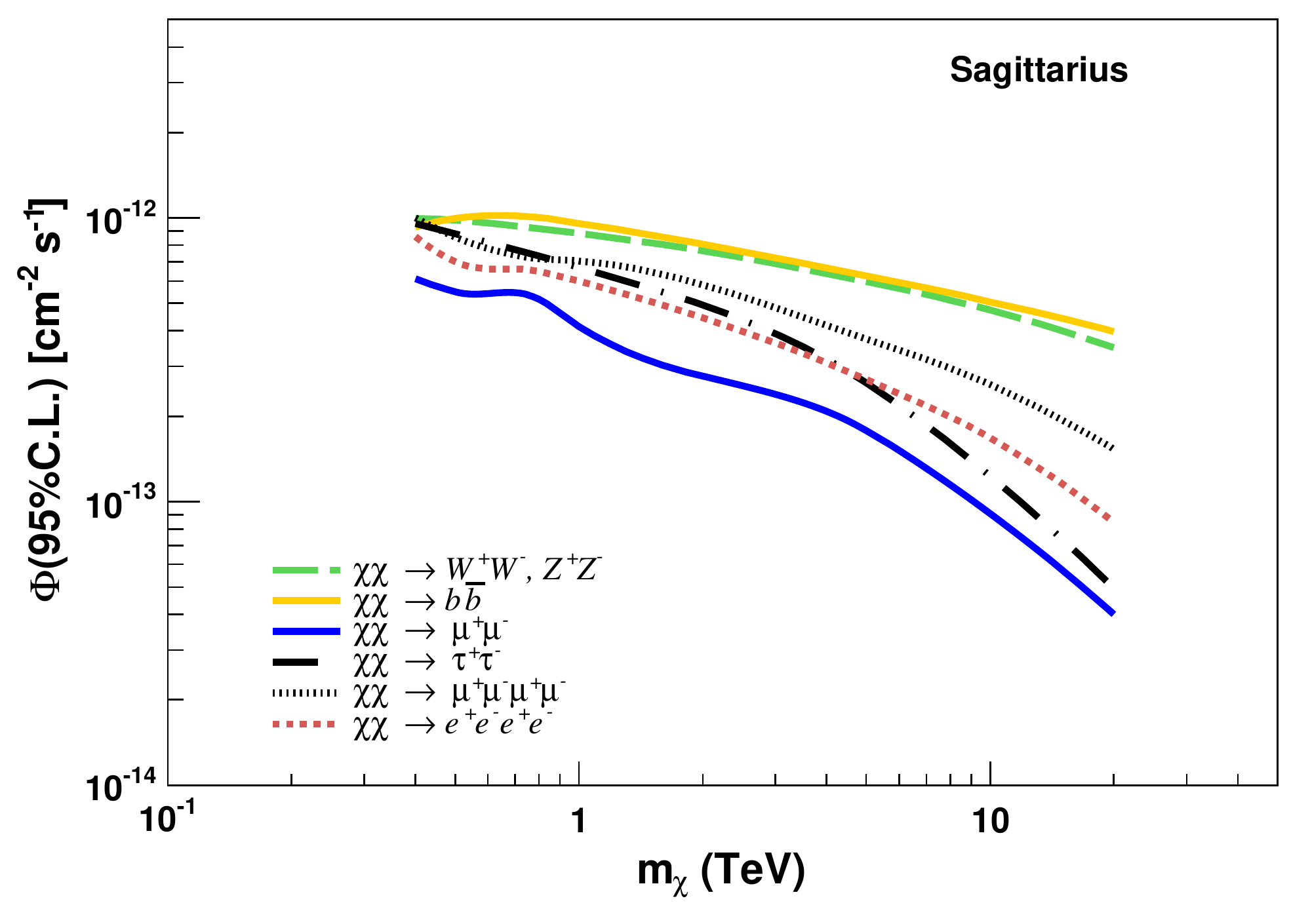}
  \caption{Upper limits at 95\% C.L. on the flux for Sagittarius dwarf galaxy as a function of $m_{\chi}$ and under the hypothesis of different DM particle annihilation channels, obtained with the likelihood approach.}
  \label{SgulF}
 \end{figure}


\section{Likelihood analysis}

In order to account for the specific shape of the DM induced gamma-ray spectrum and hence to improve the sensitivity search to faint signals, a maximum-likelihood analysis was developed for this study. For every energy bin $E_j = [E_{j_1},E_{j_2}]$, the number of events observed in the source region ($N_{\rm ON_j}$) and those in the background control regions ($N_{\rm OFF_j}$) follow Poisson distributions. The likelihood function for the energy bin is given by

\begin{multline}
\label{eq:like_binE}
\mathcal{L}^{\rm Poiss}(s_{j},b_{j},\alpha|N_{{\rm ON}, j},N_{{\rm OFF}, j}) =\\
\frac{{\left(s_{j} + \alpha b_{j} \right)}^{N_{{\rm ON}, j}}}{N_{{\rm ON}, j}!} e^{-(s_{j} + \alpha b_{j})} \times  \frac{{b_{j}}^{N_{{\rm OFF}, j}}}{N_{{\rm OFF}, j}!} e^{-b_{j}},
\end{multline}
\noindent where $\alpha = T_{\rm ON}/T_{\rm OFF}$ is the livetime exposure ratio and $b_{j}$ and $s_{j}$ are the background and signal estimates in the bin. The expected distribution of gamma-ray events is computed by folding the considered gamma-ray spectrum with the H.E.S.S. instrument response functions, specific to the analyzed data set. The expected signal in energy bin $E_j$, for a set of zenith- and off-axis angles ($\theta$, $\delta$), reads
\begin{multline}
  s_{j} = T_{\rm obs}\int\limits_{E_{j_1}}^{E_{j_2}} dE \int\limits_{0}^{\infty} dE_{t} \phi_{0} \frac{dN^\gamma(E_{t})}{dE_{t}}\\
  \times \mathcal{A}(E_{t},\theta,\delta) \times PDF(E_{t},E,\theta,\delta),
\end{multline}
\noindent where $T_{\rm obs}$ is the observatation time, $E_{t}$ is the true energy, $\phi_{0}$ is the flux normalisation, $\frac{dN^\gamma(E_{t})}{dE_{t}}$ is the assumed theoretical differential spectrum, $\mathcal{A}$ is the effective collection area and $PDF$ is the probability density function, $P(E|E_{t},\theta,\delta)$, of observing an event at the reconstructed energy $E$ for a given true energy $E_{t}$.
 
The likelihood computed over the full energy range is the product of the individual likelihood functions over all energy bins:
\begin{multline}
\label{eq:like}
\mathcal{L}^{\rm Poiss}(s,b|N_{\rm ON},N_{\rm OFF},\alpha) =\\ \prod_{j} \mathcal{L}^{\rm Poiss}_j(s_{j},\widehat{b}_{j}|N_{{\rm ON}},N_{{\rm OFF}_j},\alpha),
\end{multline}
\noindent in which $\widehat{b_{j}}$ corresponds to the best estimate of the background for a given signal $s$. 

\subsection{Flux upper limits}

For a given spectrum, the profile likelihood technique is used to assess the presence of a gamma-ray excess or to constrain the level of a possible gamma-ray emission out of the background, by determining the confidence interval of the flux normalisation. The profile likelihood function of the flux normalisation, $\phi_{0}$ reads
\begin{equation}
\lambda(\phi_{0}) = \frac{\mathcal{L}(\phi_{0},\widehat{\widehat{b}})}{\mathcal{L}(\widehat{\phi_{0}},\widehat{b})}.
\end{equation}
In the numerator, the likelihood is maximised by $\widehat{\widehat{b}}$ for fixed $\phi_{0}$, while in the denominator it is maximised globally, with $(\widehat{\phi_{0}},\widehat{b})$ corresponding to the maximum likelihood estimators.

Additionally, the results obtained have been cross-checked with the more traditional method using the upper limit on the number of gamma ray events above the energy threshold E$_{th}$, $N_{\gamma}^{95\% C.L.}(E>E_{th})$, obtained from the data analysis and listed in Table~\ref{analy}~\footnote{In this analysis, the energy threshold is defined as the energy where 10\% of the maximum effective area is reached.}. From this number, the 95\% C.L. upper limit on the integral flux can be computed for each dSph (according to equation 5 in~\cite{hessScCa}).

As a first step towards obtaining flux upper limits, the parametrization given in equation (\ref{Berg}) is considered. The corresponding upper limits on the flux for the five dSphs are shown in Figure~\ref{ulFlux} as a function of $m_{\chi}$.

In Figure~\ref{SgulF} the target source is fixed to Sgr and the flux bounds for the final state channels  $W^+W^-$ and $Z\,Z$, $b$$\overline{b}$ pairs, $\tau^+\tau^-$ pairs, $\mu^+\mu^-$ pairs and four leptons $l^+ l^- l^+ l^-$ are shown. These limits are found be as low as $\Phi_{\gamma}^{95\% C.L.}$ = 4$\times$10$^{-14}$ cm$^{-2}$ s$^{-1}$ for DM masses of ${\cal{O}}(10 \mathrm{TeV})$. Moreover, the flux limits are significantly stronger for the leptonic channels with respect to the gauge boson ones (the same applies to the Higgs/quark channels, which behave fairly similarly to the $W^+W^-$ and $Z\,Z$ ones) while amongst the leptonic channels, the four muon final state flux limit is less stringent than the corresponding two-muon final state.

\subsection{Exclusion limits on the WIMP self-annihilation cross-section.}
\label{subsec:likelihood}

In order to constrain the DM annihilation cross-section, the procedure described in \citep{Ackermann:2011aa,Ackermann:2013yva} is followed, and the uncertainty on the $J$-factor is incorporated as a nuisance parameter in the profile likelihood of each target. For each dSph $i$, the distribution of its $J$-factor can be described by a logNormal distribution, the mode, $\overline{\log_{10}(J_i)}$, and standard deviation, $\sigma_i$, of which are reported in Table~\ref{tab:dwarfs}.

The flux upper limits together with the astrophysical factors can be used to constrain the WIMP self-annihilation cross-section in a model specific context. The confidence intervals on the velocity-weighted annihilation cross-section $\sv$ as a function of the WIMP particle mass and for a given halo profile are obtained by constructing the profile likelihood function
\begin{equation}
\lambda(\sv) = \frac{\mathcal{L}(\sv,\widehat{\widehat{J}},\widehat{\widehat{b}})}{\mathcal{L}(\widehat{\sv},\widehat{J},\widehat{b})},
\end{equation}
\noindent where the nuisance parameters are the $J$-factor. $J$ and the background rate, $b$. $\widehat{\widehat{J}}$ and $\widehat{\widehat{b}}$ denote the $J$-factor and background rate which maximize the likelihood computed at $\sv$, whereas $(\widehat{\sv},\widehat{J},\widehat{b})$ is the triplet of velocity-weighted annihilation cross-section, $J$-factor and background estimate values which globally maximize the likelihood function. In the analysis, the upper limits on the annihilation cross-section are obtained under the restriction of a null and positive value of the parameter $\sv$, so that the obtained limits are conservative when a deficit of events is observed in the source region.

The profile likelihood described above allows for a straightforward combination of the results obtained from several targets. Under the assumption that the DM characteristics are shared by all targets, the combined likelihood for an assembly of  dSph is simply the product of the individual likelihood of all dSph.

\begin{figure*}[t]
  \centering
  \includegraphics[width=0.45\textwidth]{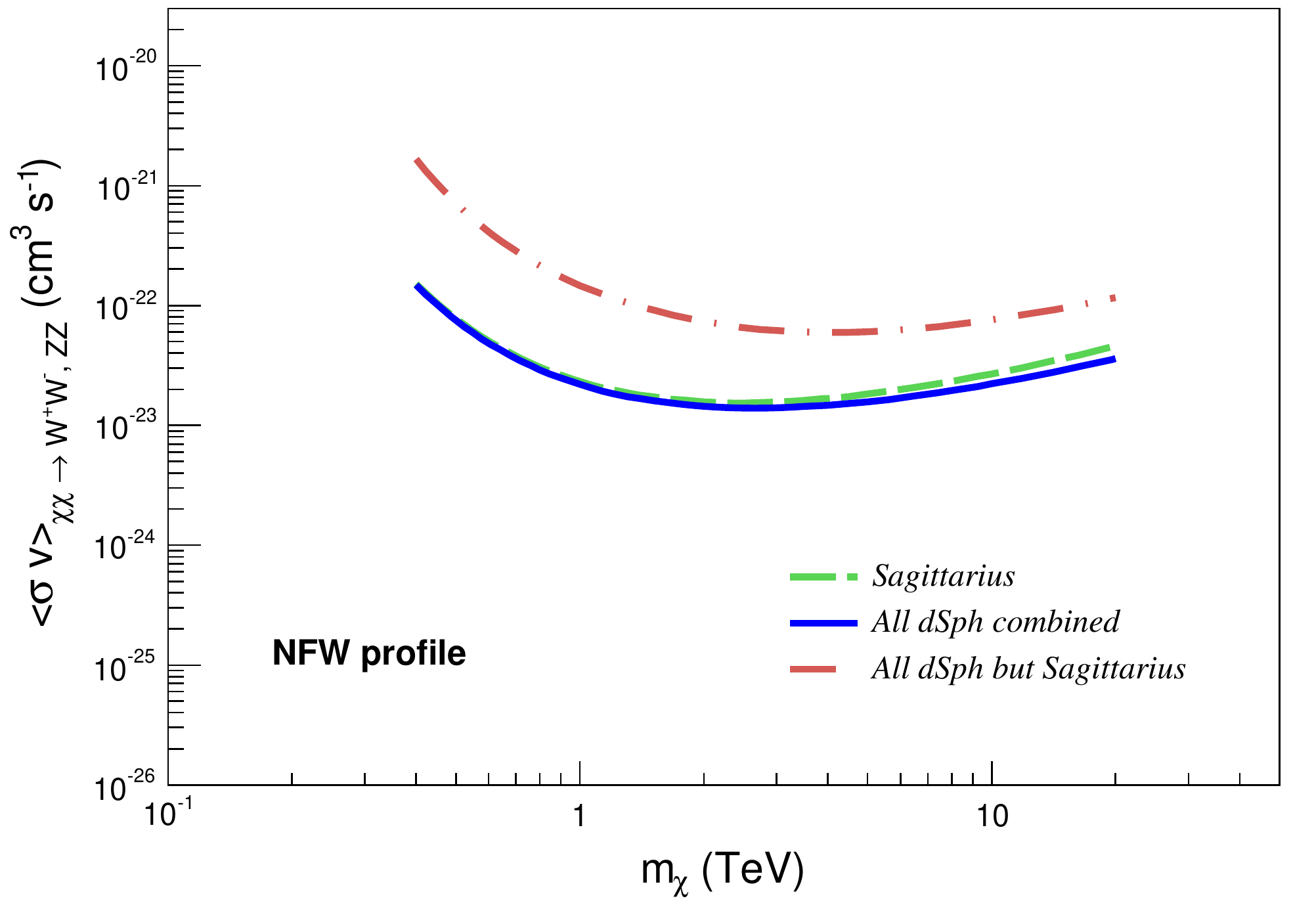}
  \includegraphics[width=0.45\textwidth]{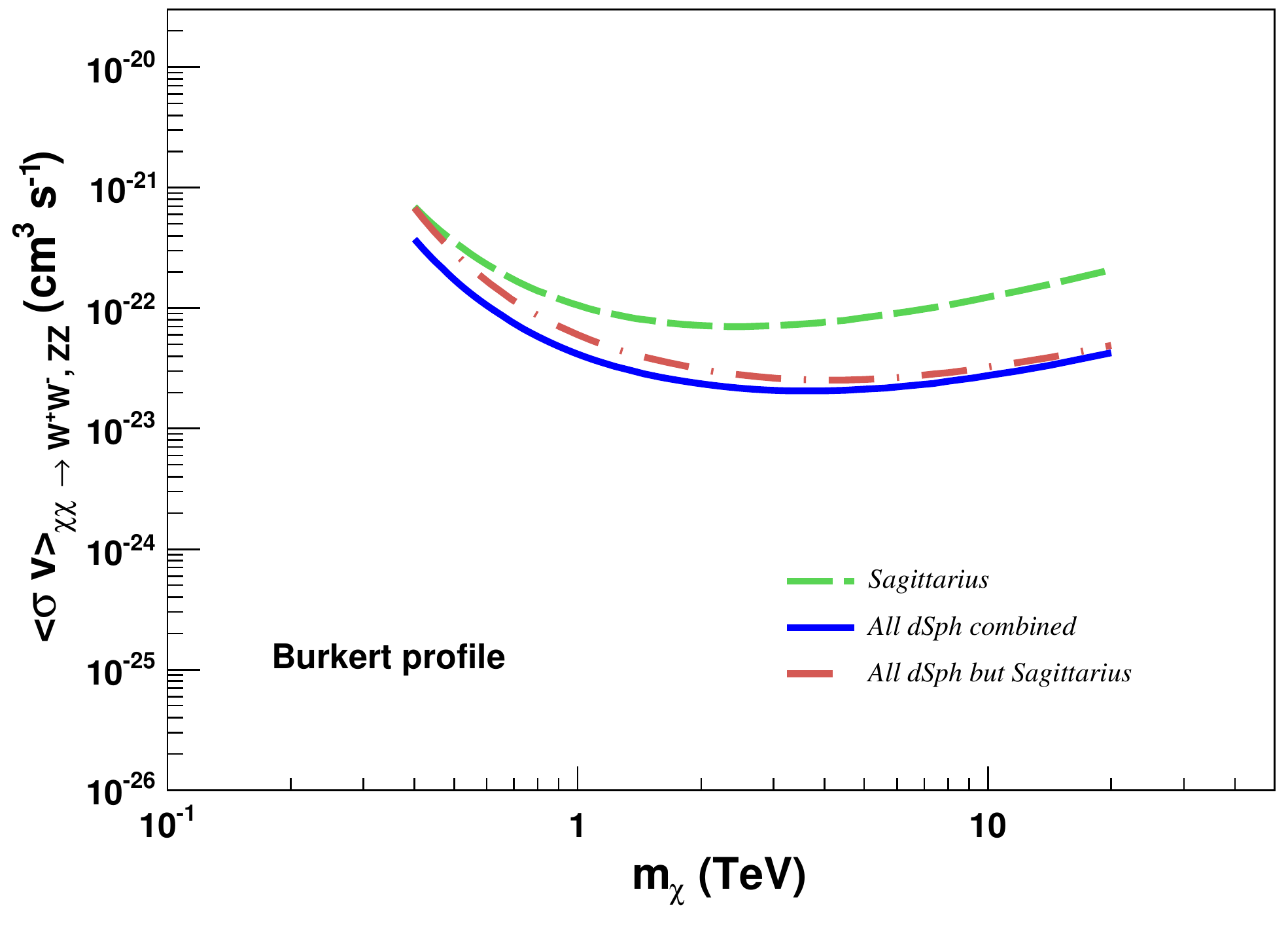}
  \caption{Combined exclusion limits at 95\% C.L. on the velocity-weighted WIMP self-annihilation cross-section versus the DM particle mass $m_{\chi}$, under the hypothesis of DM particle annihilation in the $W^{+}W^{-}$ and $Z\,Z$ final states as parametrized in Eq.~(\ref{Berg}) and for the two hypotheses of NFW and Burkert halo profiles. The results refer to the combination of all five dwarf galaxies examined in this work and the combination of all but Sgr. The Sgr bounds are also shown for comparison.}
  \label{SigmaV3}
 \end{figure*}

The combined analysis has been performed on all five dSphs. The combinations including or not the Sgr data are shown separately in Figure~\ref{SigmaV3} to illustrate the impact of this last target on the final result.

\begin{figure*}[t]
  \centering
  \includegraphics[width=0.45\textwidth]{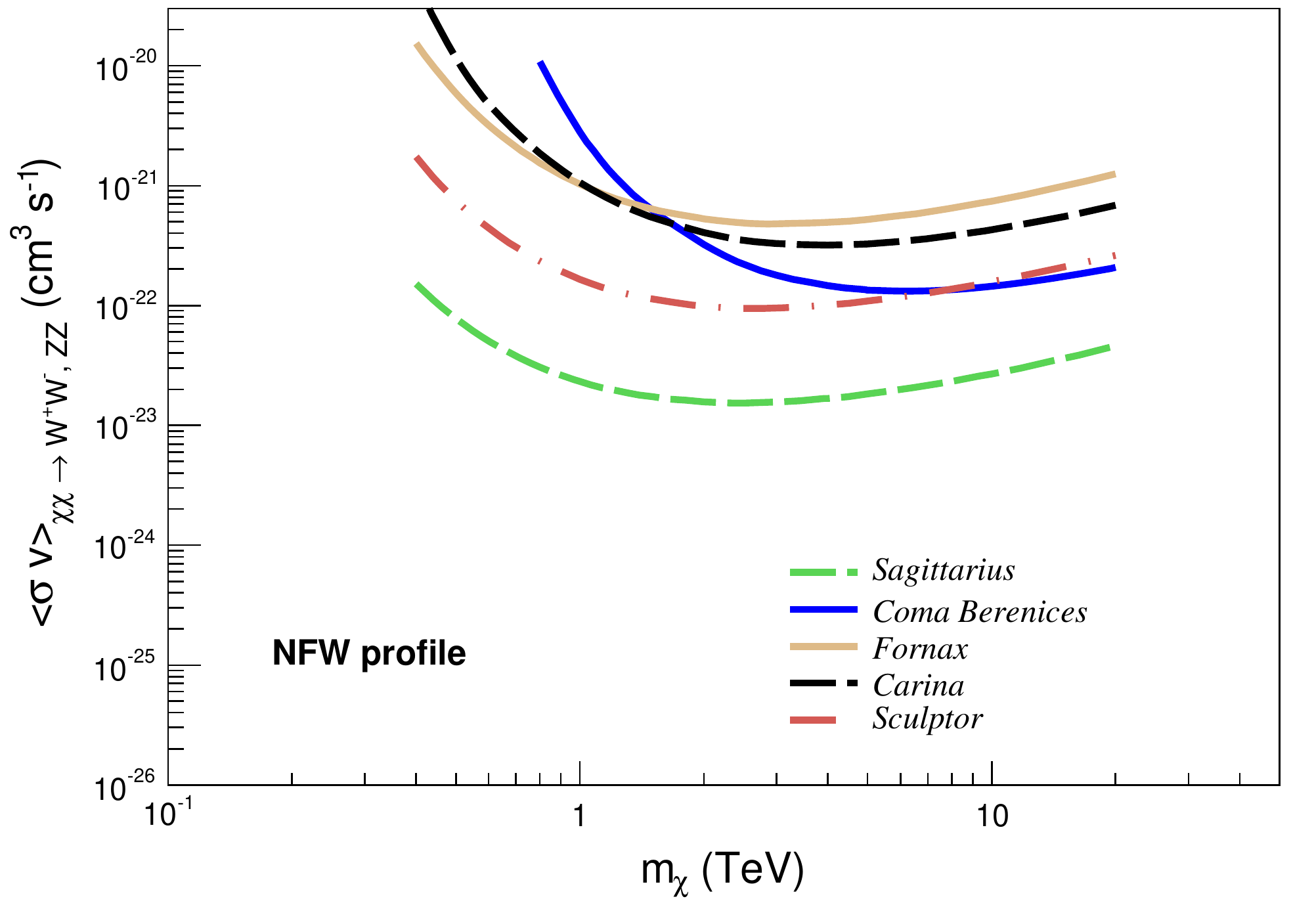}
  \includegraphics[width=0.45\textwidth]{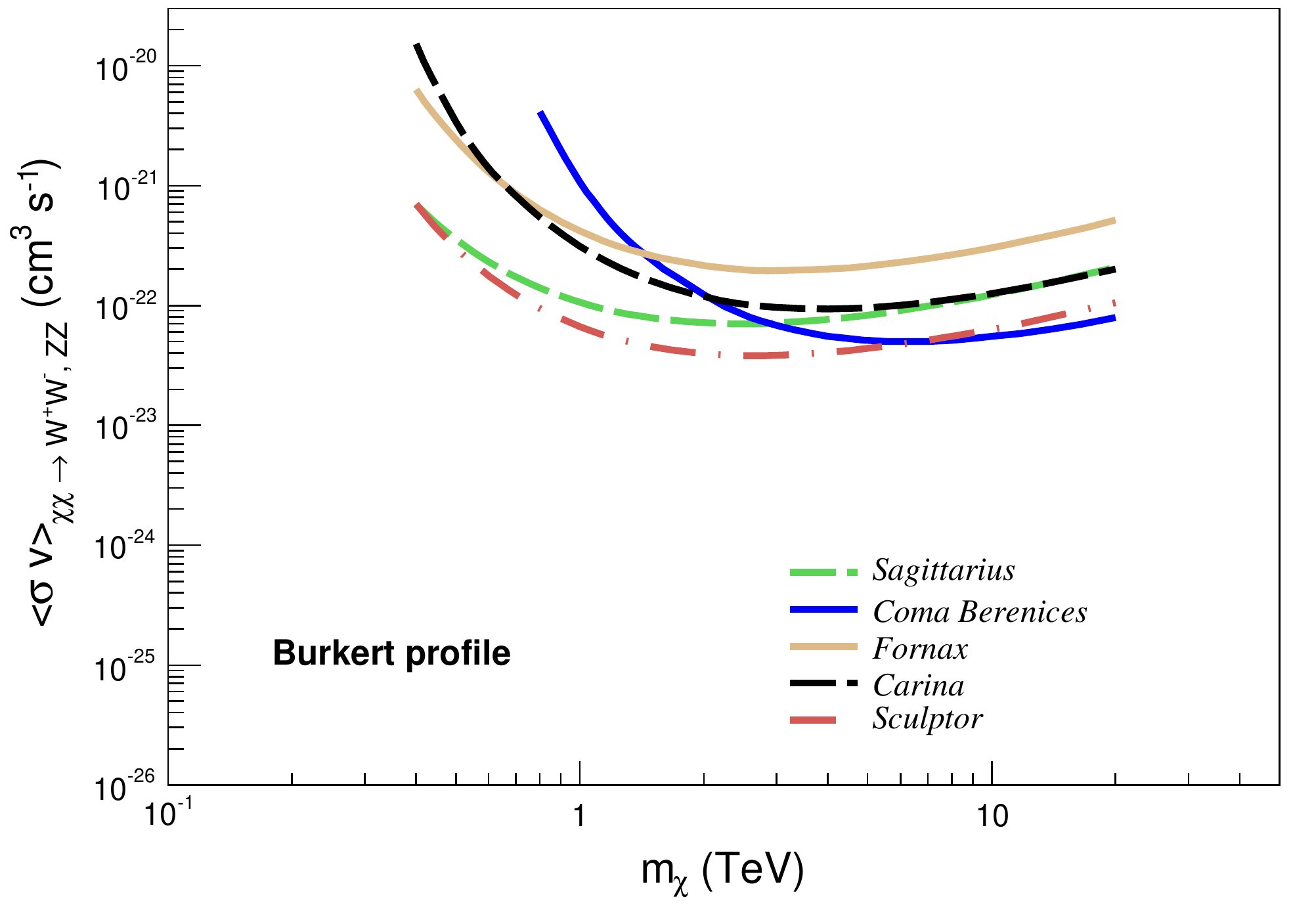}
  \caption{Exclusion limit at 95\% C.L. on the velocity-weighted WIMP self-annihilation cross-section versus the DM particle mass $m_{\chi}$, under the hypothesis of DM particle annihilation in the $W^{+}W^{-}$ and $Z\,Z$ final states as parametrized in Eq.~(\ref{Berg}) and for the two hypotheses of NFW and Burkert halo profiles.}
  \label{SigmaV3A}
 \end{figure*}

In Figure~\ref{SigmaV3A} the exclusion curves are presented for the five target dwarf galaxies separately and assuming a DM particle annihilating into $W^+W^-$ and $Z\,Z$ final states, parametrized according to Eq.~(\ref{Berg}). For all dSphs two halo profiles have been explored through the values of the $J$-factors reported in Table~\ref{tab:dwarfs} showing the effect of the astrophysical uncertainties. In the case of a NFW DM profile, it is found that the strongest constraint comes from Sgr, since it is characterized by the largest exposure, large $J$-factor and relatively low threshold. The exclusion limits depend on the particle mass and the best sensitivity is reached around 2~TeV with the value of $\sv,\sim$ 1.6$\times$10$^{-23}$ cm$^3$ s$^{-1}$ with Sgr. 
    
The combination of the various targets is shown in Figure~\ref{SigmaV3}. For a NFW profile, the combined result is only marginally improved (with a minimum value of $\sv,\sim$ 1.4$\times$10$^{-23}$ cm$^3$ s$^{-1}$) compared to Sgr bounds. 

\begin{figure}[t]
  \centering \includegraphics[width=0.45\textwidth]{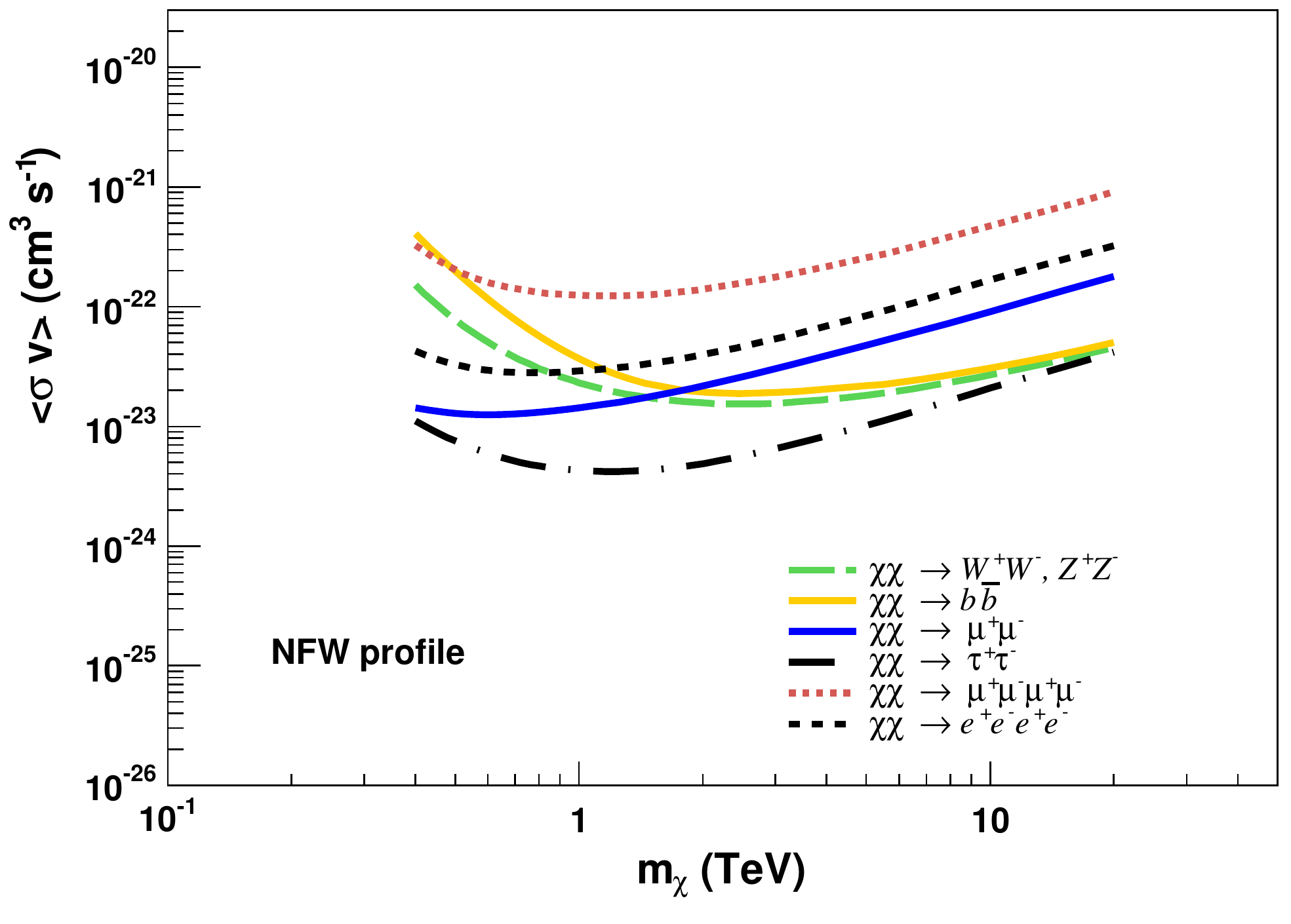}
  \caption{Exclusion limit from the Sgr on the velocity-weighted annihilation cross-section versus the DM particle mass $m_{\chi}$ and under the hypothesis of DM particle annihilation in different channels.}
  \label{SigmaSG}
\end{figure}

These exclusion curves are subject to uncertainties also in the particle physics side. In order to illustrate the particle physics uncertainties, Figure~\ref{SigmaSG} shows the limits obtained from the Sgr alone for different annihilation channels, assuming the NFW halo profile as reference. The strongest bounds are obtained for annihilation into a $\tau^+ \tau^-$ final state. For heavy masses, bounds obtained for gauge boson final state channel become competitive. This is consistent with the qualitative picture previously described: the $\tau^+ \tau^-$ channel has a relatively hard spectrum combined with a substantial normalization in the photon yield, so that even at values of the DM mass not too far above the experimental threshold the constraints are sizable. However, for very large $m_{\chi}$ most of the ``soft'' gamma rays associated to the gauge boson channel fall above threshold, where $\mathcal{A}$ is sufficiently large, and the constraint on this channel becomes comparably stronger. Similarly, one can interpret the two muon final state channel constraints: although weaker than the $\tau^+ \tau^-$ at high mass, since the photon yield is lower, it becomes comparable at low mass. Note that this channel, which has the hardest spectrum amongst the considered channels, provides the best sensitivity for small DM particle masses. Compared to that, the four-lepton final states are less constrained because of the low branching ratios into gamma rays. Note also how the peak in sensitivity is pushed to higher values with respect to the two muon channel for kinematical reasons. The four-electron final state is more constrained than the four-muon one since electrons tend to radiate much more than muons due to their smaller mass. The corresponding results on the $b\bar{b}$ channel are comparable to the $W^+ W^-$ ones. 

\section{Discussion}

\subsection{Comparison to supersymmetric scenarios}

In order to compare the H.E.S.S. exclusion limits to the predictions of a realistic particle physics model, scans were performed over the parameter space of the Next-to-Minimal Supersymmetric Standard Model (NMSSM)~\citep{Maniatis:2009re,Ellwanger:2009dp}. The NMSSM is the simplest extension of the Minimal Supersymmetric Standard Model (MSSM) that can address a number of phenomenological and theoretical issues of the latter, by the sole addition of a singlet chiral superfield. It has been shown~\citep{Cao:2012fz,King:2012tr} that the NMSSM can reproduce the observed Higgs boson mass in a wider region of the supersymmetric parameter space, without having to resort to very special configurations for the parameter values, offering a wider spectrum of possibilities for DM phenomenology as well. Note also that it reduces to the MSSM when the mass of the singlet goes to infinity, so that it includes MSSM DM candidates as subcases, notably constrained MSSM benchmark points often found in the literature.  

A scan of the general NMSSM has been performed, with parameters defined at the supersymmetric scale, looking in particular for relatively heavy neutralinos that fall within the H.E.S.S. sensitivity region but without resorting to extremal parameter values that would be theoretically unmotivated~\footnote{ The parameter ranges that have been considered are the following (all masses in GeV): $0.1 < \lambda, \kappa<1$; $-1500<A_\lambda, A_\kappa<-300$; $400 (700)<\mu, M_1, M_2(, M_3)<2000 (3000)$; $1000<m_{L,2}, m_{L,3}, m_{e,2}, m_{e,3}, m_{Q,2}, m_{Q,3}, m_{u,2}, m_{u,3}, m_{d,2}, m_{d,3}<3000$; $(-)1800<A_t(, A_l)<(-)3000$; $1.5<\tan\beta<60$.}. 

The parameter space has been scanned using the latest micrOMEGAs 3.2 code~\citep{Belanger:2013oya} that is linked to the NMSSMTools 4.0 package~\citep{Ellwanger:2005dv,Das:2011dg,Muhlleitner:2003vg}. The accepted points satisfy all relevant theoretical and experimental constraints
~\citep{Maniatis:2009re,Ellwanger:2009dp}. In particular, the lightest CP-even Higgs boson mass is taken to be compatible with LHC observations, although with a slightly looser mass bound (between $121$ GeV and $130$ GeV). For the DM relic density, the range 0.089 $<$ $\Omega_{\rm DM}h^2$ $<$ 0.14 is considered, which envelops the recent Planck measurements \cite{Ade:2013zuv} while allowing for a sufficiently efficient parameter space scan.

\begin{figure}[b]
  \centering
  \includegraphics[width=0.46\textwidth]{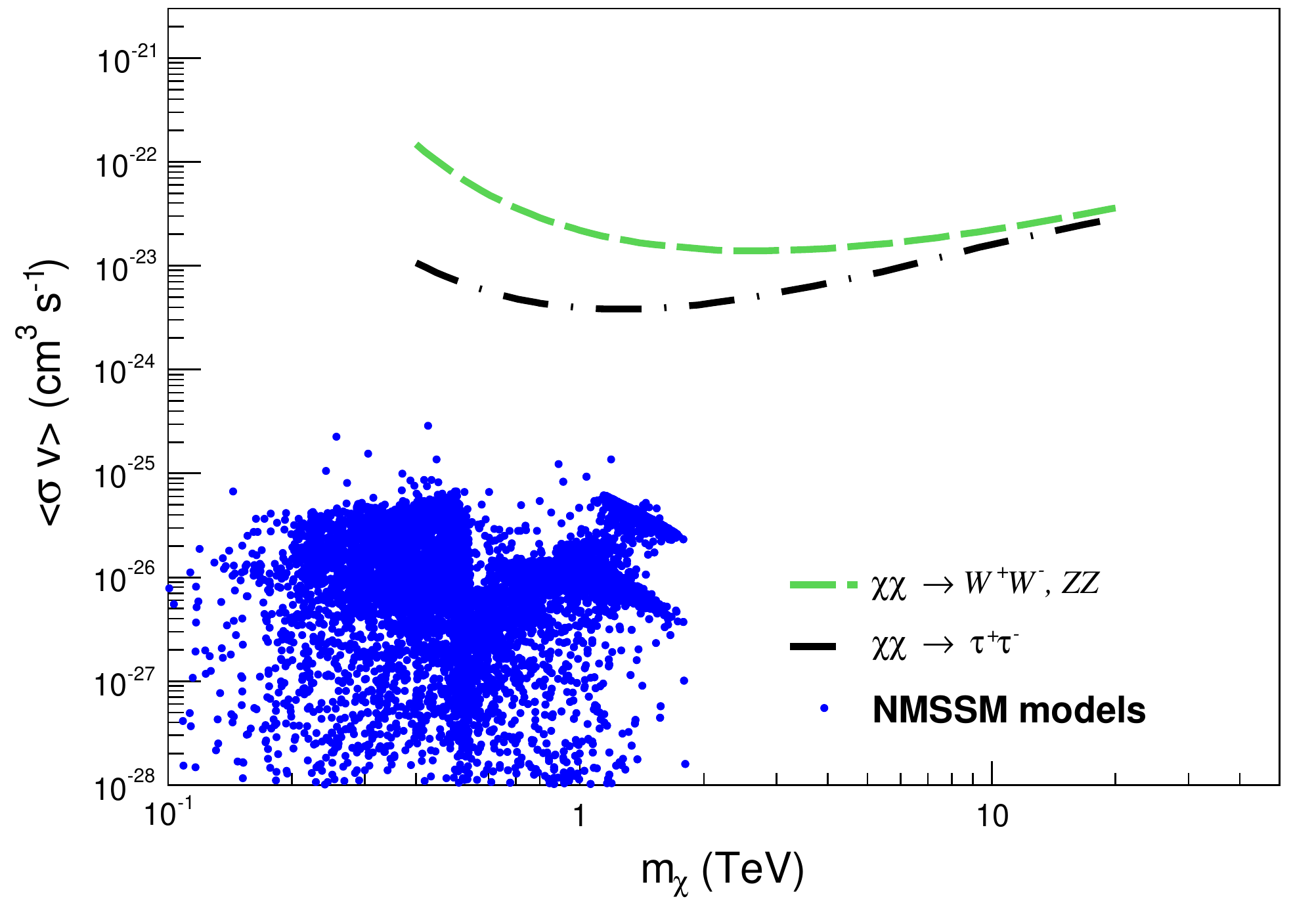}
  \caption{Exclusion limits on the velocity-weighted annihilation cross-section versus the DM particle mass. The limits combine the results from the five dwarf galaxies assuming a NFW DM density profile and two WIMP annihilation final states: $W^+W^-$, $Z\,Z$ and $\tau^+\tau^-$ channels. NMSSM models scan is also shown (blue markers).}
  \label{fig:limit}
\end{figure}

The results of the combined dSph analysis for hadronic channels, most relevant in supersymmetric DM models, are presented in Figure~\ref{fig:limit} for two WIMP annihilation final states ($W^+W^-$/$Z\,Z$ and $\tau^+\tau^-$). Along with the obtained exclusion curves, the $(m_{\chi}, \sv)$ values obtained from the NMSSM scans (blue points) are also presented.  The results show that the H.E.S.S. best exclusion limit from the dSphs combined analysis is reached at about 1 TeV with the value of $\sim$ 3.9$\times$10$^{-24}$ cm$^3$ s$^{-1}$. 
The obtained limits complement the constraints established by Fermi for lower masses using a similar approach~\cite{Ackermann:2013yva}. These limits are amongst the best obtained so far from IACT observations of dSph targets. They are close to those obtained recently with a deeper observation of Segue 1 dSph conducted by MAGIC~\cite{magicsegue}, when the $J$-factor uncertainties are also considered.
Our results are not sensitive enough to constrain typical supersymmetric scenarios, as in such setups the neutralino self-annihilation cross-section typically lies around the benchmark value of $3\times 10^{-26}$ cm$^3$ s$^{-1}$.
Note that the NMSSM scans do not account for the possibility of non-perturbative (Sommerfeld) enhancement of the neutralino self-annihilation cross-section, usually studied in scenarios where the neutralino is mostly wino and not necessarily a thermal relic DM candidate.
The inclusion of such effects would enhance $\sv$, with a strong dependence on $m_\chi$. Given the rather ``singular'' and by now severely constrained nature of such setups, along with the large number of different supersymmetric models that predict such configurations, it is preferable to stick to more representative supersymmetric scenarios. Sommerfeld-enhanced scenarios will nonetheless be discussed in more detail in section \ref{sec:Sommerfeld}.

Note also that using the most optimistic values of the $J$-factors reported in Table~\ref{tab:dwarfs} could enhance the obtained limits by an order of magnitude, and bring the exclusion bounds closer to the highest cross-section values obtained in the proposed NMSSM scan. The higher sensitivity of Cherenkov systems expected with the observations conducted by means of the H.E.S.S. fifth large telescope and even more with the advent of the new generation Cherenkov Telescope Array (CTA) observatory~\cite{Doro:2012xx}, could constrain supersymmetric models that are currently inaccessible to any ground- or space-based experiment, including the LHC.

\subsection{Comparison to models with Sommerfeld enhancement}\label{sec:Sommerfeld}
During the last few years, interesting new electron and positron cosmic-ray data have been released: notably, the PAMELA experiment reported an anomalous rise in the positron fraction spectrum~\citep{Adriani:2008zr}, independently confirmed later by Fermi-LAT~\citep{FermiLAT:2011ab}, more recently by AMS-02~\citep{Aguilar:2013qda}, and complemented by the ATIC~\citep{Chang:2008aa}, Fermi-LAT~\citep{Abdo:2009zk,Ackermann:2010ij} and H.E.S.S.~\citep{Aharonian:2008aa,Aharonian:2009ah} measurements of the total $e^+ + e^-$ flux. Despite the presence of several competing astrophysical explanations (see~\cite{Serpico:2011wg} for a review), a number of models appeared in the literature that predict very large DM self-annihilation cross-sections by invoking non-perturbative effects that are inefficient in the primordial universe but can become extremely efficient at present times, close to the zero velocity limit (Sommerfeld enhancement). These models typically also need to be ``leptophilic'' to avoid other constraints, such as excessive antiproton production.

The new H.E.S.S. exclusion bounds are compared to the model-independent best-fit regions presented in~\cite{Cirelli:2008pk}. These regions are, for different final states, adjusted to PAMELA and AMS-02 positron fraction spectrum, and independently to the total $e^+ + e^-$ flux measured by Fermi-LAT and H.E.S.S. Given the large number of Sommerfeld-enhanced models that have appeared in the literature, instead of comparing the H.E.S.S. exclusion bounds to specific theoretical models, it is chosen to compare them to these representative values of masses and cross-sections, anyway targeted by most of the relevant model-building.
 
\begin{figure}[h]
  \centering
  \includegraphics[width=0.47\textwidth]{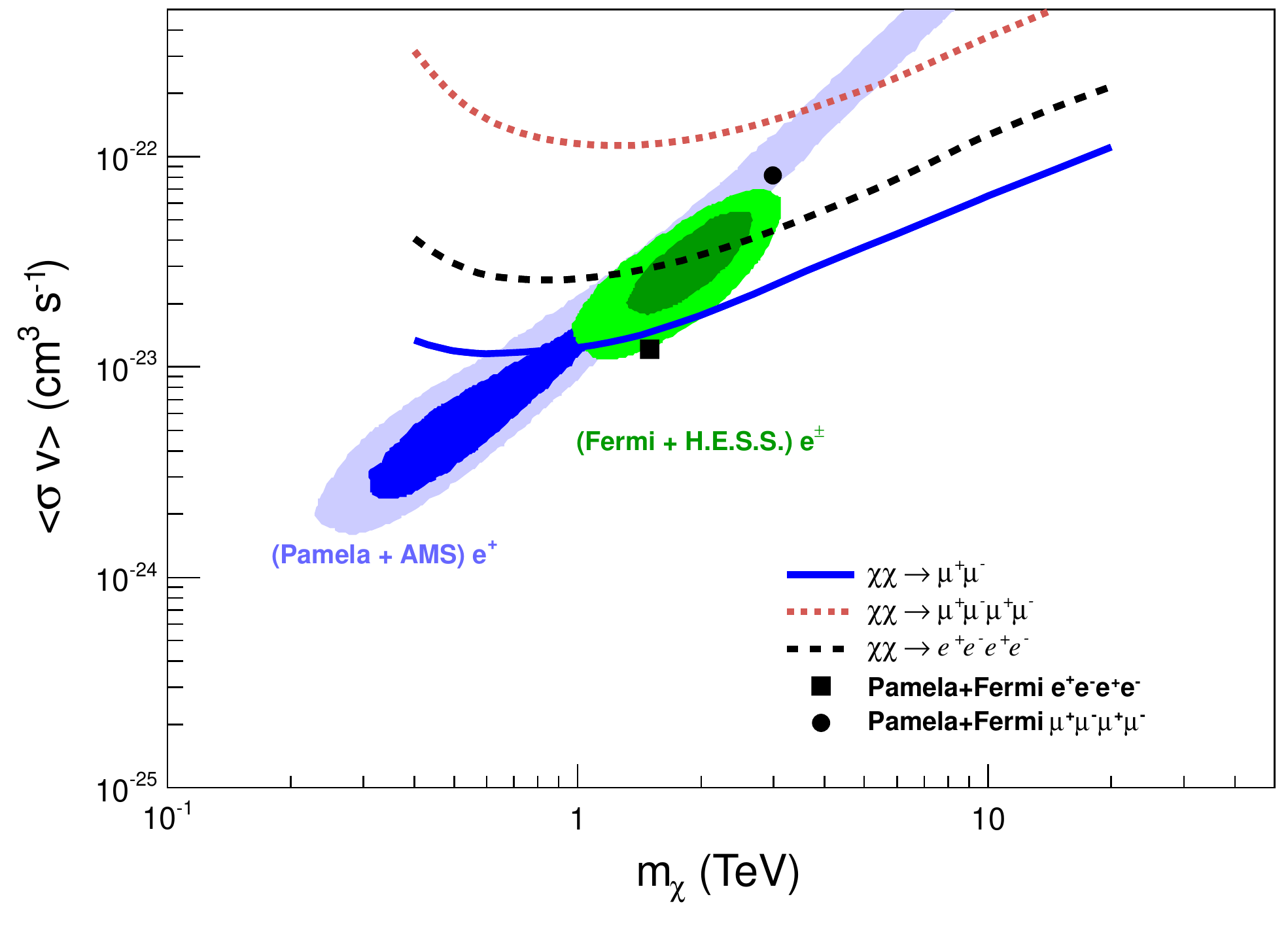}
  \caption{Exclusion limits on the velocity-weighted annihilation cross-section versus the DM particle mass. The limits combine the results from the five dwarf galaxies observed with H.E.S.S. and assume a NFW DM density profile. The results are compared with regions of the $(m_{\chi}, \sv)$ plane favored by AMS and PAMELA measurements of the positron fraction (blue contours) and by Fermi-LAT and H.E.S.S. measurements of the electron and positron fluxes (green contours), both at 3$\sigma$ and 5$\sigma$, assuming annihilation into a $\mu^+\mu^-$ final state. The corresponding results for the $e^+e^-e^+e^-$ and $\mu^+\mu^-\mu^+\mu^-$ channels are also shown.}
  \label{fig:limit2}
\end{figure}

The results of the combined dSphs analysis for leptonic channels are presented in Figure~\ref{fig:limit2}. This figure compares the obtained limits to the $(m_{\chi}, \sv)$ values fitting the AMS and PAMELA measurements of the positron fraction (blue contours) and the Fermi-LAT and H.E.S.S. measurements of the electron and positron fluxes (green contours), assuming annihilation into a $\mu^+\mu^-$ final state. The corresponding best-fit points assuming an $e^+e^-e^+e^-$ or a $\mu^+\mu^-\mu^+\mu^-$ final state are also shown. In the case of the two-muon channel, the new H.E.S.S. limits exclude the most interesting part of the remaining parameter space, in particular the regions reconciling the AMS/PAMELA/Fermi-LAT/H.E.S.S. observations. Moreover,  the four-lepton best-fit points are at the verge of exclusion, being less than a factor $2$ away from the bounds obtained in this work. Although not explicitly shown here, the H.E.S.S. upper limits also confirm the exclusion bounds obtained by the PAMELA, AMS and Fermi-LAT collaborations assuming a $\tau^+ \tau^-$ annihilation channel. It is worth to note once more that also in this scenario, an increase in sensitivity with future Cherenkov systems will enable a definitive independent test for the DM interpretation of the positron excess.

 
\section{Summary}
\label{sec:summary}
During the last years, five dwarf spheroidal galaxies have been observed with H.E.S.S. for more than 140 hours in the framework of the search for TeV gamma-ray emission from annihilation of DM particles. In the absence of any (individual or combined analysis) significant signal, constraints on the annihilation cross-section as a function of the DM mass are derived. These limits have been obtained for a combined analysis of five dwarf spheroidal galaxies, and by directly taking into account the uncertainty of the DM distribution, which is an innovative procedure in VHE gamma-ray astrophysics. The new limits have been compared to theoretical scenarios compatible with experimental results coming from the LHC experiments. Even if the obtained bounds are at present relatively far from thermal relic benchmark values, such measurements are extremely important since they can probe DM mass values lying beyond the reach of other experiments. Moreover, when it comes to positron excess-motivated leptophilic DM scenarios, the new H.E.S.S. limits already contribute to constrain the DM interpretations of the lepton spectral features observed by a series of experiments. The next generation of IACTs will further improve these limits and enable the scrutining of a larger variety of DM scenarios.


\begin{acknowledgments}
The support of the Namibian authorities and of the University of Namibia
in facilitating the construction and operation of H.E.S.S. is gratefully
acknowledged, as is the support by the German Ministry for Education and
Research (BMBF), the Max Planck Society, the French Ministry for Research,
the CNRS-IN2P3 and the Astroparticle Interdisciplinary Programme of the
CNRS, the U.K. Science and Technology Facilities Council (STFC),
the IPNP of the Charles University, the Czech Science Foundation, the Polish 
Ministry of Science and  Higher Education, the South African Department of
Science and Technology and National Research Foundation, and by the
University of Namibia. We appreciate the excellent work of the technical
support staff in Berlin, Durham, Hamburg, Heidelberg, Palaiseau, Paris,
Saclay, and in Namibia in the construction and operation of the
equipment. 

The authors acknowledge the useful collaboration of Gregory Martinez for the computation
of the astrophysical density factors. We are grateful to Marco Cirelli
for  providing the electron/positron best-fit regions referred to in the conclusive discussion.  

This work has been supported by the \textit{Investissements d'avenir}, Labex \textit{ENIGMASS}.
 
\end{acknowledgments}

\bibliography{HESS_Dwarf_2}
  
\end{document}